\documentclass[reprint]{revtex4-2}

\usepackage{graphicx,mathdots}
\usepackage{dcolumn}
\usepackage{bm,float,physics,bbold,amssymb}
\usepackage{hyperref}
\hypersetup{
  colorlinks   = true,    
  urlcolor     = blue,   
  linkcolor    = blue,   
  citecolor    = blue 
}
\usepackage[mathlines]{lineno}

\usepackage{xcolor}
\usepackage{soul}
\usepackage{lipsum}
\usepackage{mathtools}
\usepackage{cuted}

\begin{document}

\title{Utilizing discrete variable representations for decoherence-accurate numerical simulation of superconducting circuits}

\author{Brittany Richman$^{1,2,3}$}
\email{brr215@umd.edu}
\author{C. J. Lobb$^{1,2,4}$}
\author{Jacob M. Taylor$^{1,3,5}$}
\affiliation{$^1$Joint Quantum Institute (JQI), College Park, Maryland 20742, USA\\
$^2$Department of Physics, University of Maryland, College Park, Maryland 20742, USA\\
$^3$Joint Center for Quantum Information and Computer Science (QuICS), College Park, Maryland 20742, USA\\
$^4$Maryland Quantum Materials Center (QMC), College Park, Maryland 20742, USA\\
$^5$National Institute of Standards and Technology (NIST), Gaithersburg, Maryland 20899, USA}

\date{\today}

\begin{abstract}

Given the prevalence of superconducting platforms for uses in quantum computing and quantum sensing, the simulation of quantum superconducting circuits has become increasingly important for identifying system characteristics and modeling their relevant dynamics. Various numerical tools and software packages have been developed with this purpose in mind, typically utilizing the harmonic oscillator basis or the charge basis to represent a Hamiltonian. In this work, we instead consider the use of discrete variable representations (DVRs) to model superconducting circuits. In particular, we use `sinc DVRs' of both charge number and phase to approximate the eigenenergies of several prototypical examples, exploring their use and effectiveness in the numerical analysis of superconducting circuits. We find that not only are these DVRs capable of achieving decoherence-accurate simulation, i.e., accuracy at the resolution of experiments subject to decay, decoherence, and dephasing, they also demonstrate improvements in efficiency with smaller basis sizes and better convergence over standard approaches, showing that DVRs are an advantageous alternative for representing superconducting circuits.

\end{abstract}

\maketitle

\section{Introduction}
\label{intro}

Quantum superconducting circuits have emerged as one of the most promising platforms for building quantum computers, thanks to their scalability, tunability, and compatibility with existing fabrication technologies. A key aspect of developing this technology is the development of computational tools to aid in the design, modeling, and fabrication of various components, akin to the software tool-chains utilized for semiconductor circuits. With this in mind, there have been various efforts~\cite{klots, Gely_2020, Groszkowski2021scqubitspython, qiskitmetal, DiPaolo2021, PhysRevB.103.174501, Minev2021, heinsoo_2021_4944797, Chitta_2022, Aumann_2022, Rajabzadeh2023analysisofarbitrary} to develop accessible software tools and refine modeling methods to achieve efficient and accurate numerical simulation of these systems.

Of particular relevance is what we colloquially refer to as `decoherence-accurate simulation,' i.e., simulation that surpasses a given level of accuracy achievable in experiment, below which resolution is limited by errors such as systemic parameter variability, decay, decoherence, and dephasing. This is conceptually analogous to the principle of `chemical accuracy'~\cite{CA1,CA2} from chemical physics, which defines the target accuracy for numerical models based on the anticipated experimental accuracy. In our case of interest --- superconducting qubits --- we generally assume some degree of decay, decoherence, and dephasing of the superconducting levels. If we characterize this rate as that associated with the decoherence time~$T_2$~\cite{companion,QuEng} of the system, we can estimate that energetic effects relevant for high fidelity operations may no longer matter for energy scales much less than $1/T_2$. This provides a definition of `decoherence-accurate simulation' --- the accuracy required from our numerical simulations to be on par with  experimental accuracy, determined by the decoherence in a system. In what follows, we consider a decoherence accuracy of $10^{-6}$ GHz ($T_2 \approx 1 \text{ ms}$), consistent with the best superconducting qubits in use today~\cite{fluxcoherence,fluxcoherence2,Place2021,transmoncoh}.

With this definition in hand, we have a quantifiable threshold to assess not only when a numerical simulation is sufficiently accurate, but also at what point an analysis has become less efficient for computational purposes. Any simulation that achieves accuracy beyond this threshold is necessarily expending more computational resources for an accuracy level not verifiable in experiment. However, as sources of error are managed, minimized, or eliminated, we are motivated to continue to improve the numerical accuracy of simulations.

This accuracy level hinges on the basis that we choose to represent the Hamiltonian, as well as the truncation associated with this (usually infinite) basis set. Most numerical tools~\cite{klots, Groszkowski2021scqubitspython, Aumann_2022, Rajabzadeh2023analysisofarbitrary} utilize some combination of the harmonic oscillator basis and the charge basis~\cite{devoret,devsum} with a sufficiently large enough truncation to ensure proper convergence of energy levels. Here, we consider the use of discrete variable representations (DVRs)~\cite{DVRvargensinc,genDVR1,genDVR2}, specifically various `sinc DVRs'~\cite{chemphyssinc,chemphysgensinc,DVRvargensinc,periodicDVR}, which generalize and expand upon the charge basis to offer a variety of choices in basis and truncation. DVRs are basis sets that discretize one of a system's degrees of freedom according to a particular grid, where the individual basis states are localized about the grid points on which they are defined. They employ the diagonal approximation to represent the potential and any other operator which is a function of the discretized variable, resulting in trivial diagonal forms for these Hamiltonian contributions.

As a consequence of the errors introduced by the diagonal approximation, DVRs are inherently non-variational in nature, sometimes resulting in level convergence below the reference or true energy. This property and its associated pitfalls are well-documented~\cite{DVRvar1,DVRvar2,DVRvar3,DVRvar4,DVRvargensinc}, as DVRs in various forms have long been used in chemical and nuclear physics~\cite{chemphys1,chemphys2,chemphys3,chemphys4,chemphys5,chemnucphys1,chemnucphys2,chemphyssinc,chemphysgensinc}. Additionally, Hamiltonian contributions from operators that are \textit{not} functions of the discretized variable, while capable of being analytically evaluated, often result in nontrivial, full matrices. In spite of this, we show that we can achieve and surpass decoherence accuracy in three prototypical superconducting circuits: the LC oscillator, the fluxonium, and the transmon. We demonstrate in these examples that the DVRs we consider outperform the typical sparse method, the finite difference method, and in many cases match or outperform traditional numerical approaches.

This paper is organized as follows. Section~\ref{S2} provides an overview of the DVRs employed in our numerical analysis, including specific functional forms, variable definitions, and operator representations. Numerical results are presented in Section~\ref{S3}, highlighting the features of these DVRs in the LC oscillator, the fluxonium, and the transmon, where we compare their performance to standard approaches. We conclude and discuss the implications of our results in Section~\ref{outlook}. In the appendices that follow, we provide an overview and derivation of DVRs as well as additional numerical details for our analysis.

\section{Overview of sinc DVRs}
\label{S2}

Here we provide an overview of the various DVRs we consider in Section~\ref{S3}, including the explicit forms of the basis states and relevant operator representations for numerical analysis. In general, DVRs are formed by projecting a discrete grid of points onto some orthonormal and complete basis set, and demanding that the resulting states themselves be orthonormal and complete. Relevant details and derivations for the generic construction of a DVR can be found in Appendix~\ref{DVRbkgrd}. In particular, we focus on sinc DVRs, which are created using the Fourier basis.

Sinc DVRs leverage both the orthonormality and completeness of the Fourier basis on various intervals as well as the Fourier transform, which connects the conjugate variables of a given system.
With our focus being superconducting circuits, we follow the usual circuit quantization procedures~\cite{devoret}, taking the Cooper pair number $N$ and superconducting phase of a node $\theta$ as our conjugate variables. These dimensionless variables are related to the usual node charge $Q$ and node flux $\Phi$ of a circuit via $N = Q/(-2e)$ and $\theta = 2 \pi \Phi / \Phi_0$, where $e$ is the elementary electron charge and $\Phi_0$ is the magnetic flux quantum. As a consequence, the basis states for each of the following DVRs are all expressed in terms of $N$ and~$\theta$.

\subsection{The traditional sinc DVR}
\label{tradsinc}

The traditional sinc DVR is constructed by projecting a set of discretized points onto the continuous Fourier basis, taking advantage of the Fourier series expansion to enable the creation of a DVR. This permits a discretization of either $N$ or $\theta$, while its respective conjugate variable, $\theta$ or $N$, is continuous and bounded. See Appendix~\ref{tradsincDER} for the details of its construction.

\begin{figure}
\centering
\includegraphics[width= 0.48\textwidth]{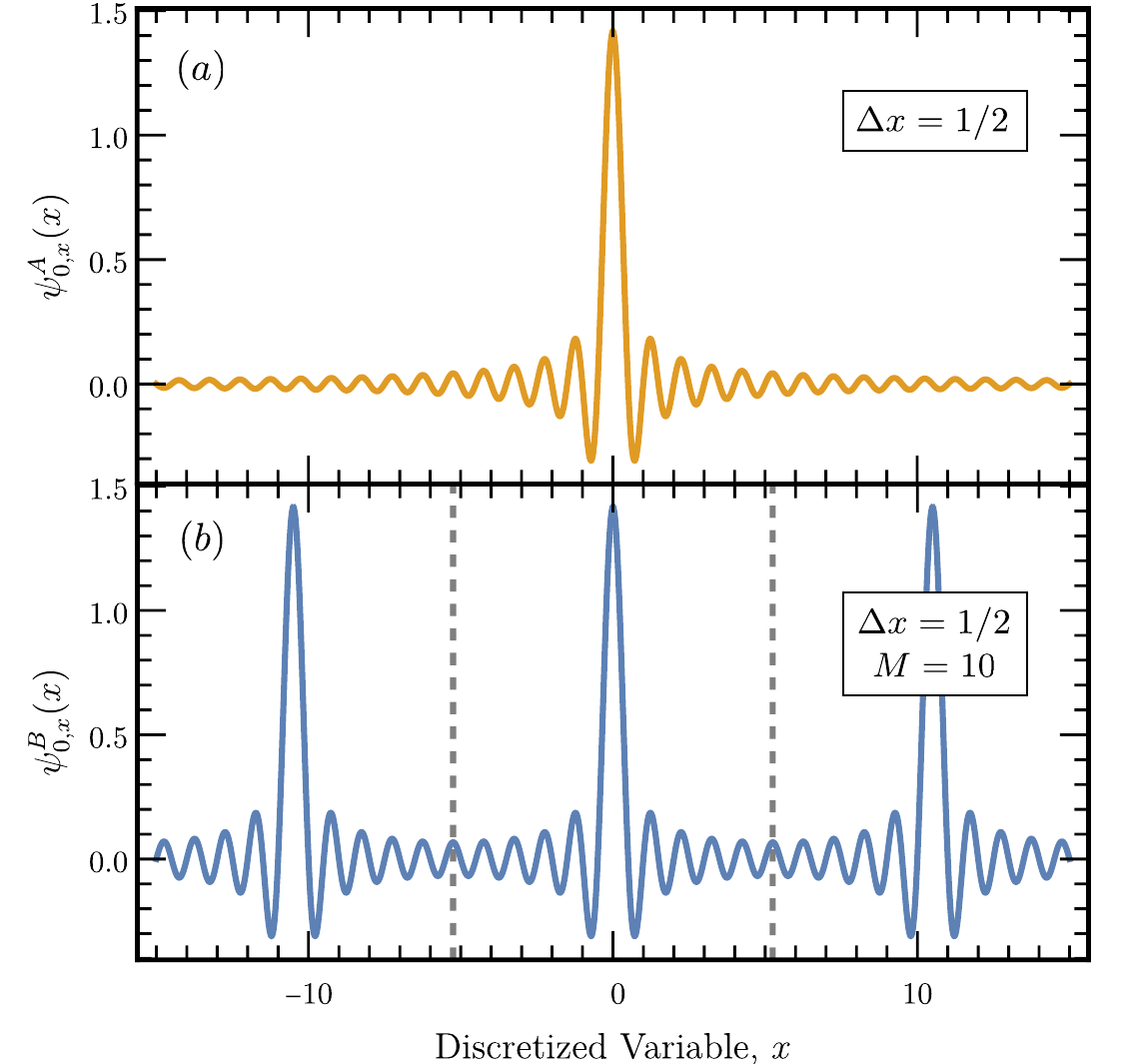}
\caption{Functional forms of the DVR basis states for the (a) traditional sinc DVR and (b) truncated sinc DVR, both for an indexed $\alpha = 0$ state. Remaining basis states are generated by translating these functions in the positive and negative directions. The functional width of these basis functions is increased or decreased by increasing or decreasing the grid size. In (b), we have included dashed grey vertical lines to indicate the bounded region on which the DVR is valid, representing the interval on which its basis functions are periodic. Variations in both the number of states $M$ and the grid size $\Delta x$ affect this interval.}
\label{fig:sinc}
\end{figure}

Discretizing $\theta$ yields a phase DVR with the basis states
\begin{equation}
\label{eq:tradsincTH}
\ket{\psi_{\alpha,\theta}^{A}} = \frac{1}{\sqrt{2 N_{\rm max}}} \int_{-N_{\rm max}}^{N_{\rm max}} dN \ e^{i N \theta_\alpha} \ket{f_N} \enspace ,
\end{equation}
which in functional form are sinc functions, 
\begin{equation}
\label{eq:tradsincTHF}
\psi^{A}_{\alpha,\theta} (\theta) = \sqrt{\frac{N_{\rm max}}{\pi}} \frac{\sin \left[ N_{\rm max} (\theta - \theta_\alpha)  \right] }{N_{\rm max} (\theta - \theta_\alpha)} \enspace,
\end{equation}
hence the naming of this DVR as a `sinc DVR.' We index the states with an integer $\alpha$ corresponding to their grid point, the discretized variable (here $\theta$), as well as an additional superscript $A$ to distinguish this DVR from the truncated sinc DVR discussed in the following subsection. This DVR is valid for an infinite, uniform grid of points $\theta_\alpha = \alpha \Delta \theta$, spaced according to $\Delta \theta = \pi / N_{\rm max}$, where $\pm N_{\rm max}$ bounds the continuous variable $N$. The states $\ket{f_N}$ represent the $\theta$-space functions used in the projector to construct the DVR, i.e., the continuous Fourier basis, where $\bra{\theta}\ket{f_N} = f(\theta,N) = e^{-i N \theta}/\sqrt{2 \pi}$~[analogous to Eq.~(\ref{eq:CFB})]. Note that in this case, the DVR's basis functions are non-periodic with infinite extent in $\theta$-space (as shown in Fig.~\ref{fig:sinc}a), but periodic in conjugate space (charge number) on the interval $[-N_{\rm max},N_{\rm max}]$, since the Fourier transform of $\psi_{\alpha,\theta}^{A} (\theta)$ is simply the band-limited complex exponential $e^{-i N \theta_\alpha}/\sqrt{2 N_{\rm max}}$.

Similarly, discretization of $N$ yields a charge number DVR with the basis states
\begin{equation}
\label{eq:tradsincN}
\ket{\psi_{\alpha,N}^{A}} = \frac{1}{\sqrt{2 \theta_{\rm max}}} \int_{-\theta_{\rm max}}^{\theta_{\rm max}} d\theta \ e^{-i N_\alpha \theta} \ket{f_\theta} \enspace ,
\end{equation}
which in functional form are again sinc functions, but now in $N$-space:
\begin{equation}
\psi_{\alpha,N}^{A} (N) = \sqrt{\frac{\theta_{\rm max}}{\pi}} \frac{\sin \left[ \theta_{\rm max} (N - N_\alpha)  \right] }{\theta_{\rm max} (N - N_\alpha)} \enspace.
\end{equation}
In this case, we require an infinite, uniform grid $N_\alpha = \alpha \Delta N$, spaced according to $\Delta N = \pi / \theta_{\rm max}$, where $\pm \theta_{\rm max}$ bounds the continuous variable $\theta$. The states $\ket{f_\theta}$ represent the $N$-space functions used in the projector to construct the DVR, $\bra{N}\ket{f_\theta} = f(\theta,N)^* =  e^{i N \theta}/\sqrt{2 \pi}$~[analogous to Eq.~(\ref{eq:CFB})]. Here, we find that this DVR's basis functions are instead non-periodic with infinite extent in $N$-space, but periodic in phase space on the interval $[-\theta_{\rm max},\theta_{\rm max}]$.

While these DVRs are quite general, with a variety of available discretization choices, we note that one of the standard basis sets used in superconducting circuit analysis, the charge basis~\cite{devoret,devsum}, is in fact a specific instance of the charge number DVR in Eq.~(\ref{eq:tradsincN}). The charge basis is typically represented with the charge number and phase states,  
\begin{equation}
\ket{N} = \frac{1}{\sqrt{2 \pi}} \int_{-\pi}^{\pi}  e^{-i \theta N} \ket{\theta} d\theta
\end{equation}
and
\begin{equation}
\ket{\theta} = \frac{1}{\sqrt{2 \pi}} \sum_{N = -\infty}^{\infty} e^{ i \theta N} \ket{N} \enspace ,
\end{equation}
respectively, where charge number is integer-valued and phase is $2 \pi$-periodic and continuous. This is exactly the charge number DVR in Eq.~(\ref{eq:tradsincN}), with the specific choice of $\theta_{\rm max} = \pi$ and $\Delta N =1$. The widespread implementation of this discretization choice, i.e., the charge basis, originates from the simplest superconducting circuit, a Josephson junction connecting two islands of superconducting material. This system is described by the Hamiltonian
\begin{equation}
\hat{H} = 4 E_C \hat{N}^2 - E_J \cos \hat{\theta} \enspace ,
\end{equation}
with $E_C = e^2/2 C_J$ the junction's charging energy, $E_J$ its Josephson energy, and $C_J$ the junction capacitance. Critically, this Hamiltonian's potential is $2 \pi$-periodic, thus demanding that charge be quantized and integer-valued. However, this periodicity is not ubiquitous, enabling the more general use of both charge number and phase DVRs, as we explore in Section~\ref{S3}.

\subsection{The truncated sinc DVR}
\label{truncsinc}

The traditional sinc DVR requires an infinite number of basis states, which must necessarily be truncated in any numerical representation. Numerical truncation introduces a source of error, particularly for state spaces not large enough to be considered `approximately infinite.' However, we can similarly construct sinc DVRs that are instead finite-sized by definition using the partially discrete Fourier basis. As in the traditional case, we can discretize either the charge number or phase variable, yielding DVRs we interpret as truncations of the traditional sinc DVRs already discussed. This truncated DVR instead exploits the discrete Fourier transform for its creation --- details of the construction of this DVR can be found in Appendix~\ref{truncsincDER}. We now highlight their form and relevant features, along with their connection to the DVRs already presented.

A truncated sinc DVR that discretizes phase is given by the basis states
\begin{equation}
\label{eq:truncsincTH}
\ket{\psi_{\alpha,\theta}^{B}} = \frac{1}{\sqrt{2 M +1}} \sum_{n = -M}^{M} e^{i N_n \theta_\alpha} \ket{f_n} \enspace ,
\end{equation}
and a finite-sized grid $\theta_\alpha = \alpha \Delta \theta$, with $\alpha$ an integer such that $\alpha = -M, ..., M$, and $\Delta \theta \Delta N = 2 \pi/ (2M+1)$. As before, we index the states according to their grid point $\alpha$, the discretized variable (here $\theta$), as well as an additional index $B$ to distinguish this DVR from the traditional sinc DVR discussed in the previous subsection. The basis used to construct this DVR is the partially discrete Fourier basis in $\theta$-space, $\bra{\theta} \ket{f_n} = f_n(\theta) = e^{-i N_n \theta} / \sqrt{2 \theta_{\rm max}}$ [analogous to Eq.~(\ref{eq:xcon})], where charge number is already discretized according to $N_n = n \Delta N$ and phase is continuous and bounded by $\theta_{\rm max} = \pi/\Delta N$. These states are represented in $\theta$-space with the functional form,
\begin{equation}
\label{eq:truncsincTHF}
\begin{split}
\psi_{\alpha,\theta}^{B} (\theta) &= \frac{1}{(2M+1)\sqrt{\Delta \theta}} \sum_{n = -M}^{M} e^{-i N_n(\theta-\theta_\alpha)} \\
&= \frac{1}{(2M+1)\sqrt{\Delta \theta}} \frac{\sin \left[ \frac{\pi}{\Delta \theta}  (\theta- \theta_\alpha)  \right]}{\sin \left[ \frac{\pi}{\Delta \theta}  (\theta - \theta_\alpha) / (2M+1) \right]}
\end{split} \enspace ,
\end{equation}
where we see that now, not only is the basis for this DVR finite-sized, its states are periodic in the discretized variable on the interval $[-\theta_{\rm max},\theta_{\rm max}]$, as shown in Fig.~\ref{fig:sinc}b.

Analogously, a truncated sinc DVR that discretizes charge number takes the form
\begin{equation}
\label{eq:truncsincN}
\ket{\psi_{\alpha,N}^{B}} = \frac{1}{\sqrt{2 M +1}} \sum_{n = -M}^{M} e^{-i N_\alpha \theta_n} \ket{g_n} \enspace ,
\end{equation}
utilizing a finite-sized grid of points $N_\alpha = \alpha \Delta N$ with $\alpha = -M, ..., M$, which are again spaced according to $\Delta \theta \Delta N = 2 \pi/ (2M+1)$. In this case, the basis used to construct the DVR is the partially discrete Fourier basis in $N$-space, $\bra{N} \ket{g_n} = g_n(N) = e^{i N \theta_n}/\sqrt{2 N_{\rm max}}$ [analogous to Eq.~(\ref{eq:pcon})], where phase is already discretized, $\theta_n = n \Delta \theta$, and charge number is continuous and bounded by $N_{\rm max} = \pi/\Delta \theta$. These states are represented in $N$-space with the functional form,
\begin{equation}
\begin{split}
\psi_{\alpha,N}^{B} (N) &= \frac{1}{(2M+1)\sqrt{\Delta N}} \sum_{n = -M}^{M} e^{i (N-N_\alpha)\theta_n} \\
&= \frac{1}{(2M+1)\sqrt{\Delta N}} \times \\
& \quad \quad \quad  \quad \frac{\sin \left[ \frac{\pi}{\Delta N}  (N- N_\alpha)  \right]}{\sin \left[ \frac{\pi}{\Delta N}  (N - N_\alpha) / (2M+1) \right]}
\end{split} \enspace .
\end{equation}
Once again, this DVR's basis is finite-sized and periodic in the discretized variable (now charge number), on the interval $[-N_{\rm max},N_{\rm max}]$.

These two DVRs can be interpreted as truncations of the traditional sinc DVRs in Section~\ref{tradsinc} upon comparing their respective basis states. Examining the DVRs which discretize phase [Eqs.~(\ref{eq:tradsincTH}) and (\ref{eq:truncsincTH})], we find that the traditional sinc DVR [Eq.~(\ref{eq:tradsincTH})] is a continuum limit of the truncated sinc DVR [Eq.~(\ref{eq:truncsincTH})], where we take $M \rightarrow \infty$ and keep $\Delta \theta$ fixed, resulting in a continuous and bounded conjugate variable $N$. It is then clear that the truncation of the traditional DVR's state space (and thus the discretized variable $\theta$) leads to a discretization of the continuous variable $N$, resulting in the truncated DVR states in Eq.~(\ref{eq:truncsincTH}). This description connecting the traditional and truncated DVRs applies analogously to the DVRs which discretize charge number [Eqs.~(\ref{eq:tradsincN}) and (\ref{eq:truncsincN})]. We note that this is consistent with the connection between the discrete, rotor, and continuous phase spaces described in Ref.~\cite{victor}, as one might expect, with the only distinction being that we have generalized to consider arbitrarily-sized grid spacings.

\subsection{Representing operators}
\label{OpRep}

Having established in Sections~\ref{tradsinc} and \ref{truncsinc} the various DVRs of interest, we now consider the implementation of these DVRs in numerics, specifically via the matrix representation of operators relevant for superconducting circuit analysis. Importantly, representing these operators using each DVR's basis states allows us to exploit a characteristic feature of all DVRs, namely, the `diagonal approximation.'

As outlined in Appendix~\ref{DVRbkgrd}, this approximation assumes a diagonal matrix representation for operators that are functions of the discretized variable of a DVR, essentially approximating a DVR's basis states as eigenstates of the `discretized' operator. Thus, for operators that are a function of $\theta$, a matrix representation using the traditional or traditional phase DVRs yields a diagonal form. For operators such as $\hat{\theta}$, $\hat{\theta}^2$, or $\cos(\hat{\theta} \pm 2 \pi \mathcal{A})$, found often in the Hamiltonians of superconducting circuits with Josephson junctions, we have, 
\begin{equation}
\begin{split}
\bra{\psi_{\alpha,\theta}^{(A,B)}} \hat{\theta} \ket{\psi_{\beta,\theta}^{(A,B)}} &= \theta_\alpha \delta_{\alpha \beta} \enspace , \\
\bra{\psi_{\alpha,\theta}^{(A,B)}} \hat{\theta}^2 \ket{\psi_{\beta,\theta}^{(A,B)}} &= \theta^2_\alpha \delta_{\alpha \beta} \enspace ,\\
\\
\bra{\psi_{\alpha,\theta}^{(A,B)}} \cos(\hat{\theta} \pm 2 \pi \mathcal{A}) \ket{\psi_{\beta,\theta}^{(A,B)}} &= \cos(\theta_\alpha \pm 2 \pi \mathcal{A}) \delta_{\alpha \beta}\enspace ,
\end{split} 
\end{equation}
where $\ket{\psi_{\alpha,\theta}^{(A,B)}}$ represents the traditional or truncated sinc DVRs in Eqs.~(\ref{eq:tradsincTH}) or (\ref{eq:truncsincTH}), respectively. Here we have included an often present term, $\pm 2 \pi \mathcal{A}$, which characterizes a system's external flux frustration via the ratio $\mathcal{A} = \Phi^{\rm ext}/\Phi_0$. Analogously, the traditional and truncated charge DVRs yield diagonal matrix representations for operators that are a function of $N$, such as $\hat{N}$ or $\hat{N}^2$. Their matrix elements are simply
\begin{equation}
\begin{split}
\bra{\psi_{\alpha,N}^{(A,B)}} \hat{N} \ket{\psi_{\beta,N}^{(A,B)}} &= N_\alpha \delta_{\alpha \beta} \enspace , \\
\bra{\psi_{\alpha,N}^{(A,B)}} \hat{N}^2 \ket{\psi_{\beta,N}^{(A,B)}} &= N^2_\alpha \delta_{\alpha \beta} \enspace ,
\end{split}
\end{equation}
where $\ket{\psi_{\alpha,N}^{(A,B)}}$ represents the charge number DVRs in Eqs.~(\ref{eq:tradsincN}) or (\ref{eq:truncsincN}). While this assumption greatly simplifies the numerical representation of the associated Hamiltonian terms of these operators, notably, the diagonal approximation introduces a quadrature error that is not guaranteed to be positive. Thus, DVRs are inherently non-variational basis representations, and care must be taken when examining their convergence properties.

The diagonal approximation assumes a DVR's basis states are eigenstates of the discretized variable, however, the basis set utilized in the construction of each sinc DVR we consider (the Fourier basis) holds particular relevance as well, namely, as eigenstates of the conjugate variable. For each of the phase DVRs in Eqs.~(\ref{eq:tradsincTH}) and (\ref{eq:truncsincTH}), the basis states $\ket{f_N}$ and $\ket{f_n}$ are eigenstates of $\hat{N} = i \partial / \partial \theta$, while for each of the charge number DVRs in Eqs.~(\ref{eq:tradsincN}) and (\ref{eq:truncsincN}), the basis states $\ket{f_\theta}$ and $\ket{g_n}$ are eigenstates of $\hat{\theta} = -i \partial / \partial N$. As a consequence, matrix elements of operators which are functions of the variable conjugate to the discretized variable can be evaluated analytically using the basis states of each DVR.

Considering first the traditional sinc DVRs in Section~\ref{tradsinc}, the charge number operators $\hat{N}$ and $\hat{N^2}$ are represented in the discrete phase basis [Eq.(\ref{eq:tradsincTH})] with the matrix elements,
\begin{equation}
\label{eq:Noptradsinc}
\begin{split}
\bra{\psi_{\alpha,\theta}^A} \hat{N} \ket{\psi_{\beta,\theta}^A} &= \frac{1}{2 N_{\rm max}} \int_{-N_{\rm max}}^{N_{\rm max}} dN N e^{- i N (\theta_\alpha - \theta_\beta)} \\
&= \begin{cases}
0 &\text{ if } \alpha = \beta \\
\frac{i (-1)^{\alpha + \beta}}{\Delta \theta (\alpha - \beta)} &\text{ if } \alpha \neq \beta
\end{cases} \enspace ,
\end{split}
\end{equation}
and
\begin{equation}
\label{eq:N2optradsinc}
\begin{split}
\bra{\psi_{\alpha,\theta}^A} \hat{N}^2 \ket{\psi_{\beta,\theta}^A} &= \frac{1}{2 N_{\rm max}} \int_{-N_{\rm max}}^{N_{\rm max}} dN N^2 e^{- i N (\theta_\alpha - \theta_\beta)} \\
&= \begin{cases}
\frac{N_{\rm max}^2}{3} &\text{ if } \alpha = \beta \\
\frac{2 (-1)^{\alpha + \beta}}{\Delta \theta^2 (\alpha - \beta)^2} &\text{ if } \alpha \neq \beta
\end{cases} \enspace .
\end{split}
\end{equation}
Analogously, the phase operators $\hat{\theta}$ and $\hat{\theta^2}$ are represented in the discrete charge number basis [Eq.(\ref{eq:tradsincN})] with the matrix elements, 
\begin{equation}
\label{eq:thetaoptradsinc}
\begin{split}
\bra{\psi_{\alpha,N}^A} \hat{\theta} \ket{\psi_{\beta,N}^A} &= \frac{1}{2 \theta_{\rm max}} \int_{-\theta_{\rm max}}^{\theta_{\rm max}} d\theta \ \theta e^{ i (N_\alpha - N_\beta) \theta} \\
&= \begin{cases}
0 &\text{ if } \alpha = \beta \\
\frac{-i (-1)^{\alpha + \beta}}{\Delta N (\alpha - \beta)} &\text{ if } \alpha \neq \beta
\end{cases} \enspace ,
\end{split}
\end{equation}
and
\begin{equation}
\label{eq:theta2optradsinc}
\begin{split}
\bra{\psi_{\alpha,N}^A} \hat{\theta}^2 \ket{\psi_{\beta,N}^A} &= \frac{1}{2 \theta_{\rm max}} \int_{-\theta_{\rm max}}^{\theta_{\rm max}} d\theta \ \theta^2 e^{i (N_\alpha - N_\beta) \theta} \\
&= \begin{cases}
\frac{\theta_{\rm max}^2}{3} &\text{ if } \alpha = \beta \\
\frac{2 (-1)^{\alpha + \beta}}{\Delta N^2 (\alpha - \beta)^2} &\text{ if } \alpha \neq \beta
\end{cases} \enspace .
\end{split}
\end{equation}
To find the matrix elements analogous to Eqs.~(\ref{eq:Noptradsinc})-(\ref{eq:theta2optradsinc}) assembled from the truncated sinc DVRs in Eqs.~(\ref{eq:truncsincTH}) and (\ref{eq:truncsincN}), we must evaluate finite sums rather than definite integrals, e.g.,
\begin{equation}
\label{eq:truncsincOPex}
\bra{\psi_{\alpha,\theta}^B} \hat{N} \ket{\psi_{\beta,\theta}^B} = \frac{\Delta N}{2 M +1} \sum_{n = -M}^{M} n \ e^{- i 2 \pi n (\alpha - \beta)/(2 M +1)} \enspace .
\end{equation}
While this sum and those for the other relevant operators in Eqs.~(\ref{eq:N2optradsinc})-(\ref{eq:theta2optradsinc}) can be evaluated analytically to yield closed-form expressions distinct from those already presented, these expressions are non-trivial in nature and do not provide any additional insight. Hence, we omit their specific forms here. We note that all of these matrices, while evaluated analytically, are full, a feature that is certainly not ideal for numerical analysis. Thus, we are motivated to compare the performance of these DVRs to a standard sparse method, the finite difference method. This comparison is discussed and included in our results in Section~\ref{LCoscillator}.

Yet to be discussed is an operator of particular relevance to superconducting systems --- the cosine operator represented in a discrete charge number basis. For simplicity, we restrict ourselves to the instances where $1/\Delta N$ is integer-valued, yielding the expression
\begin{equation}
\label{eq:cosoptradsinc}
\begin{split}
\bra{\psi_{\alpha,N}^{(A,B)}} \cos &( \hat{\theta} \pm 2 \pi \mathcal{A} )  \ket{\psi_{\beta,N}^{(A,B)}} \\
& = \frac{1}{2} \left( e^{\pm i 2 \pi \mathcal{A}} \delta_{\alpha, \gamma^-} + e^{\mp i 2 \pi \mathcal{A}} \delta_{\alpha, \gamma^+} \right)
\end{split} \enspace ,
\end{equation}
where $\gamma^\pm = \beta \pm 1/\Delta N$. This specific choice yields a sparse matrix with trivial values along only two bands, which is identical for both the traditional and truncated charge number DVRs. In addition, the operator's physical interpretation becomes more apparent --- as a tunneling operator between the basis states of the DVR, connecting a state at the grid point $\alpha$ to those indexed $\pm 1/ \Delta N$ away. While this matrix element can be evaluated in either the traditional or truncated case for arbitrary, non-integer $1/\Delta N$, the resulting matrices are full with non-trivial elements. Therefore, when this operator is to be evaluated, specifically in the fluxonium example in Section~\ref{Fluxonium}, we proceed with a consideration of integer-valued $1/ \Delta N$ only.

Before moving to concrete examples and implementing these matrix representations in numerical analysis, there are two important points of note. Firstly, the matrix representations utilizing the traditional sinc DVRs are technically of infinite size and must be truncated in any numerical representation. Typically, we use the expressions which assume an infinite-sized basis [Eqs.~(\ref{eq:Noptradsinc})-(\ref{eq:theta2optradsinc})], only in a matrix of some finite size. This discrepancy introduces a truncation error that decreases with increasing matrix size. In contrast, the truncated sinc DVRs we consider provide matrix representations that are already finite-sized --- their expressions are exact at any matrix size. Naturally, with large enough matrix sizes, the truncation error associated with the traditional sinc DVR is negligible, and the discrepancy between the traditional and truncated DVR cases is nonexistent. As discussed in Section~\ref{truncsinc}, as the matrix size increases, we reach a continuum limit where the truncated sinc DVR becomes the traditional one.

We also note that the finite-sized, fully discrete nature of the truncated sinc DVR necessitates that the truncated DVR's basis states [Eqs.~(\ref{eq:truncsincTH}) or (\ref{eq:truncsincN})] and the functions used to construct them ($\ket{f_n}$ or $\ket{g_n}$) are related by a discrete Fourier transform. As previously noted, these states are eigenstates of the operator conjugate to the discretized variable. Thus, this relationship enables a diagonalization of the conjugate variable (represented in the truncated sinc DVR's basis) via a discrete Fourier transform. In other words, the matrix representation of the conjugate variable in the truncated DVR's basis [e.g., Eq.~(\ref{eq:truncsincOPex})] can alternately be assembled via the discrete Fourier transform of the operator in its eigenbasis. This feature can enable more efficient construction and manipulation of these otherwise non-trivial and full matrices, as well as greatly simplify operations such as the trotterization of a Hamiltonian.

\section{Numerical features and applications of sinc DVRs}
\label{S3}

\begin{figure}
\centering
\includegraphics[width= 0.48\textwidth]{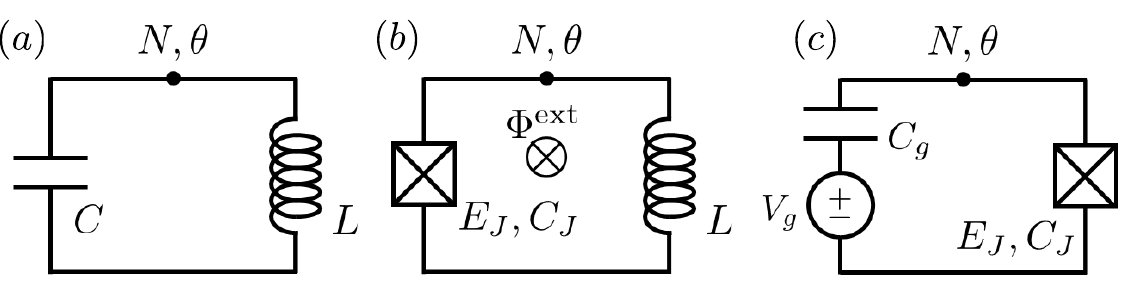}
\caption{Circuit schematics of (a) the LC oscillator, (b) the fluxonium, and (c) the transmon, the three examples considered in the text, including relevant parameter and variable definitions.}
\label{fig:examples}
\end{figure}

We now turn our attention to the focus of our investigation: the utilization of DVRs in the numerical analysis of superconducting circuits and the advantages they may provide. The DVRs outlined in Section~\ref{S2} capture a variety of possibilities with regard to boundary conditions and discretization choice, allowing for broad application to a range of systems. Here, we consider straightforward, prototypical systems [shown in Fig.~\ref{fig:examples}] to highlight each DVR's applicability, convenience, and in some cases, advantage over traditional numerical representations.

In what follows, we consider the use of these DVRs to examine three features of interest for numerical analysis: energy level convergence, basis state contributions, and phase shift implementations. We first leverage the LC oscillator and its analytical solutions as a proof-of-concept, examining the level convergence of each DVR, introducing the relevant metrics we consider to assess performance, and comparing performance to the finite difference method, which is known to be sparse. We then focus on the use of each DVR in the fluxonium as well as their more limited application in the transmon.

\subsection{LC oscillator}
\label{LCoscillator}

We now examine the use of DVRs for the evaluation of energy eigenvalues of the LC oscillator (Fig.~\ref{fig:examples}a), to serve as a point of reference to assess and understand DVR performance. The LC oscillator has exact, well-known solutions to compare to, and given this system's bounded, non-periodic potential, we can consider all four of the DVRs presented in Section~\ref{S2} for comparison. Expressed in terms of the dimensionless variables $N$ and $\theta$, the Hamiltonian is written as
\begin{equation}
\label{eq:LCham}
\hat{H} = 4 E_C \hat{N}^2 + \frac{E_L}{2} \hat{\theta}^2
\end{equation}
where the capacitive energy is given by $E_C = \frac{e^2}{2 C}$ and the inductive energy is given by $E_L = \frac{1}{L} \left(\frac{\Phi_0}{2 \pi}\right)^2$. Here the circuit's capacitance is defined as $C$, its inductance defined as $L$, with $e$ the elementary electron charge and $\Phi_0$ the magnetic flux quantum.

We assess the energy level convergence properties of each DVR by examining the absolute value of the energy difference $\Delta_n$ as a function of matrix size, i.e., the number of a DVR's basis states included in a numerical representation. Here,
\begin{equation}
\label{eq:ED}
\Delta_n = E^{\rm DVR}_n - E^{\rm ref}_n \enspace ,
\end{equation}
where $E^{\rm DVR}_n$ is the energy of $n^{\rm th}$ level found using a DVR and $E^{\rm ref}_n$ is the reference energy of that same level. In the LC oscillator, we have exact solutions to serve as a reference point: $E^{\rm ref}_n = \sqrt{8 E_C E_L} (n + 1/2)$. Naturally, as a DVR converges to the reference energy, here the exact energy of the system, $\Delta_n \rightarrow 0$.

\begin{figure}[!b]
\centering
\includegraphics[width= 0.48\textwidth]{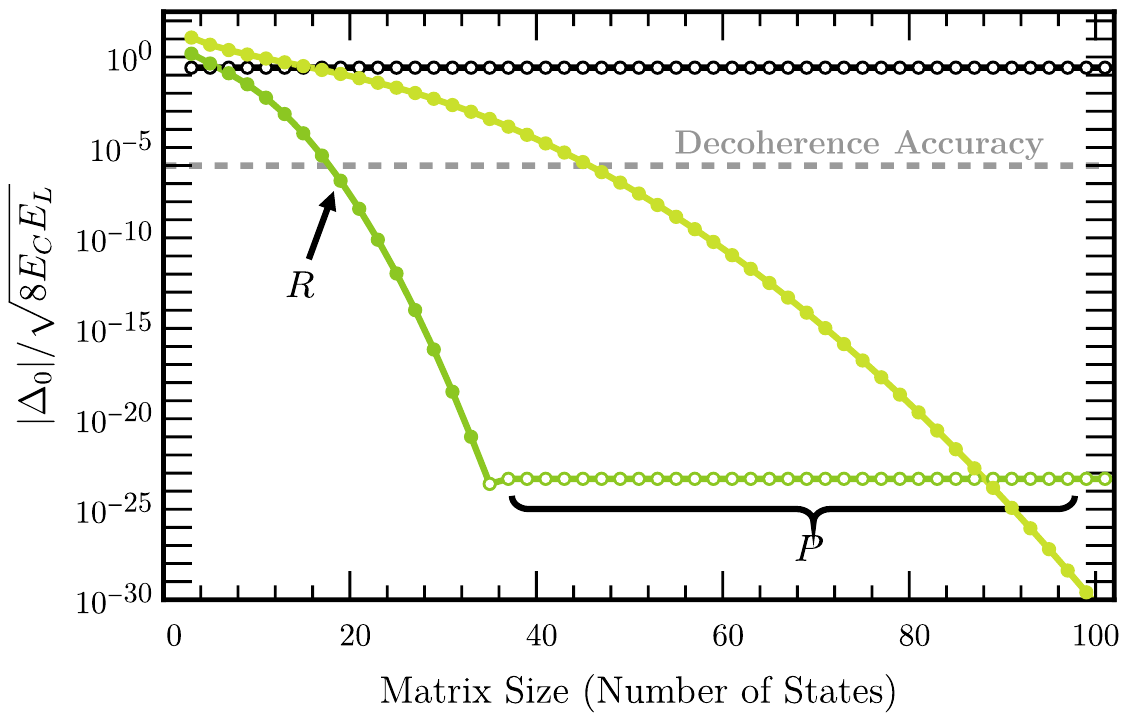}
\caption{Example behavior of the absolute value of the energy difference $|\Delta_0|$ as function of matrix size in the ground state of the LC oscillator, using the traditional charge number DVR with grid sizes $\Delta N = 3/2$ (black), $\Delta N = 1/4$ (green), and $\Delta N = 1/10$ (light green), all scaled according to the LC frequency. Decoherence accuracy (also scaled according to the LC frequency) is indicated with a horizontal gray dashed line. Additionally, the decoherence-accurate matrix size $R$ and saturation precision $P$ are identified. Open data points indicate where $\Delta_0 <0$, in order to note and highlight the non-variational nature of DVRs.}
\label{fig:curveex}
\end{figure}

In Fig.~\ref{fig:curveex}, we plot the LC oscillator's ground state level convergence as a function of matrix size for three different grid sizes in the traditional charge number DVR, corresponding to the three general categories of behavior we observe in such plots, across all DVRs and examples. For grid sizes that are too large, no amount of basis states yields an improvement in level convergence, which never approaches the reference energy or decoherence accuracy. This is indicated by the black, flat line in Fig.~\ref{fig:curveex}, where $\Delta N = 3/2$. We also find a manifestation of the non-variational nature of DVRs --- this convergence is \textit{below} the true energy, indicated by the open data points in Fig.~\ref{fig:curveex}.

For grid sizes that are too small, we observe very slow level convergence with increasing matrix sizes, which does not saturate within the working precision and/or largest matrix sizes considered, but does often reach decoherence accuracy at larger matrix sizes, indicated by the lightest descending curve in Fig.~\ref{fig:curveex}, where $\Delta N = 1/10$. The behavior of these two extremes can be understood as a consequence of the grid size's effect on the basis functions of the DVRs introduced in Section~\ref{S2}, namely, that increasing the grid size increases the function's width while decreasing the size of the conjugate variable's subspace. Essentially, grid sizes that are too large yield basis functions that are too broad to be able to resolve the wavefunction of interest, and cut off relevant, contributing regions of its conjugate subspace. On the other hand, grid sizes that are too small capture all relevant regions of conjugate space, but result in basis functions that are so narrow as to require very large numbers of states to accurately represent the function of interest.

Finally, for intermediate grid sizes, we observe level convergence that improves with increases in matrix size, surpassing decoherence accuracy and saturating with high precision to some level of accuracy relative to the reference energy, shown in Fig.~\ref{fig:curveex} for the grid size $\Delta N = 1/4$. Once again, we observe non-variational behavior, where the asymptotic convergence of the DVR is in fact below the true energy. This curve is additionally labeled with the two metrics we use to assess and compare DVR performance across the various grid sizes, DVRs, and examples. The smallest matrix size that achieves decoherence accuracy, the decoherence-accurate matrix size, is labeled with $R$, while the asymptotic behavior, or saturation precision, is labeled with $P$. Together, these metrics capture the relevant behavior of a DVR, addressing whether or not decoherence accuracy is reached, and how quickly, as well as what degree of accuracy a given DVR can achieve for a function of interest. While the curves and grid sizes shown in Fig.~\ref{fig:curveex} are specific to the LC oscillator ground state with the traditional charge number DVR, the general behavior described above extends to the other DVRs and examples we consider.

\begin{figure}
\centering
\includegraphics[width= 0.48\textwidth]{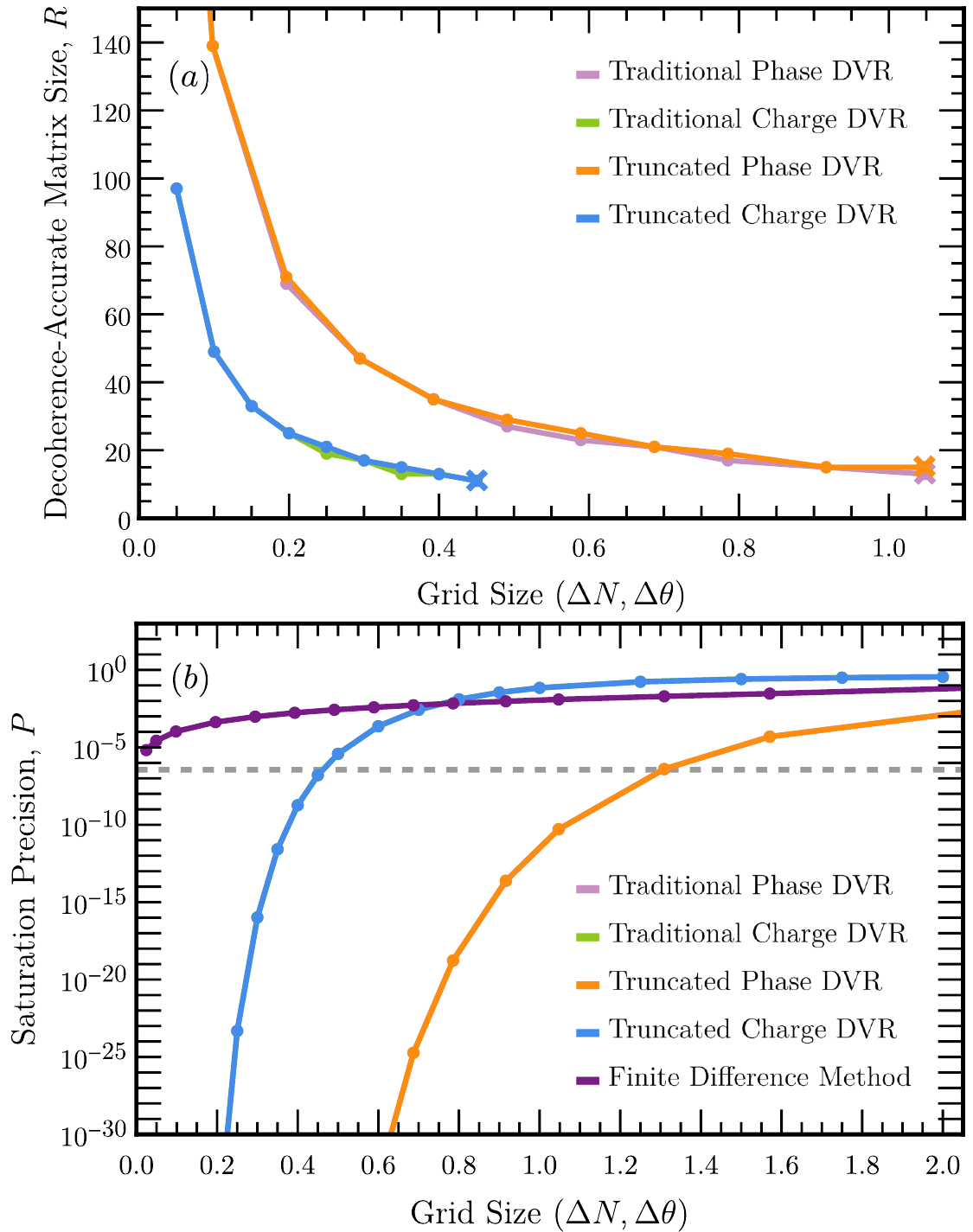}
\caption{(a) The decoherence-accurate matrix size $R$ as a function of grid size in the LC oscillator's ground state, for all four of the DVRs considered and colored according to the included legend. The grid size cutoff for $R$, i.e., the largest grid size that surpasses decoherence accuracy, is indicated with an `X' data point. (b) The saturation precision $P$ as a function of grid size in the LC oscillator's ground state, for all four of the DVRs considered as well as the lowest-order finite difference method, as indicated in the legend. Decoherence accuracy (scaled according to the LC frequency) is indicated with a horizontal gray dashed line.}
\label{fig:LCmetrics}
\end{figure}

To utilize these metrics to assess each DVR's performance in the LC oscillator, we expand upon Fig.~\ref{fig:curveex} to consider a range of grid sizes in all four DVRs. A sampling of curves akin to those in Fig.~\ref{fig:curveex} for each DVR is shown in Appendix~\ref{LCextra}, along with additional numerical details. This analysis enables an examination of both the decoherence-accurate matrix size $R$ and saturation precision $P$ as a function of the grid size, shown in Figs.~\ref{fig:LCmetrics}a and \ref{fig:LCmetrics}b, respectively.

With regard to decoherence accuracy, Fig.~\ref{fig:LCmetrics}a demonstrates an overall increase in the decoherence-acccurate matrix size $R$ as the grid size decreases, for all four DVRs. As the grid size increases, $R$ decreases, but ultimately reaches a cutoff (indicated by the final `X' data point) beyond which decoherence accuracy is never achieved. This behavior is consistent with the general understanding of the impact of grid size on a DVR's basis functions described previously. This cutoff in $R$ indicates where the grid size of a particular DVR has become too large to approximate a function of interest --- in the phase DVRs, this corresponds to grid sizes larger than $\Delta \theta = \pi/3$, while in the charge DVRs, this cutoff occurs at the grid size $\Delta N = 0.45$. As grid size decreases, we find a rapid increase in $R$, indicative of a worse performance, where the grid size is small enough to enable good approximation of the function, but requiring numerous states to do so. We also note minor variations in $R$ between the traditional and truncated DVRs of each type, particularly at smaller matrix sizes. We understand these discrepancies as a consequence of the relationship between the truncated and traditional sinc DVRs --- as the matrix size increases, the required numerical truncation in the traditional case has less impact, and we approach the continuum limit in the truncated case. However, these variations are quite small, almost negligible. Clearly, this truncation does not introduce very large error in the traditional sinc DVR and the truncated sinc DVR performs similarly well, ultimately demonstrating the effectiveness of all four DVRs to represent the system.

In Fig.~\ref{fig:LCmetrics}b, we plot the saturation precision $P$ as a function of the grid size for all four DVRs. Most notably, the traditional and truncated sinc DVRs of both types are nearly identical. As this metric corresponds to the asymptotic behavior observed at large matrix sizes, this is exactly as expected given the relationship between these DVRs --- the continuum limit is being reached. Larger grid sizes tend towards saturation at worse accuracy, while smaller grid sizes lead to rapid improvements, saturating below decoherence accuracy in numerous cases. We note that the strict condition to identify the smallest matrix size \textit{below} decoherence accuracy does not capture grid sizes such as $\Delta \theta = 5 \pi/12$, which saturate barely above decoherence accuracy.

Also visible in Fig.~\ref{fig:LCmetrics}b is an additional curve representing the performance of the finite difference method. As detailed in Appendix~\ref{FDM}, this approach discretizes the continuous phase representation of the Hamiltonian in Eq.~(\ref{eq:LCham}) over some finite region of phase space, which is determined by the chosen grid size and the number of grid points included. We consider here the usual lowest-order approximation, a three-point approximation that yields a sparse, tri-diagonal representation of the conjugate variable. Upon varying the matrix size (determined by the number of grid points) for various grid sizes, analogously to the DVRs, we examine $R$ and $P$ as a function of $\Delta \theta$. As we show in Appendix~\ref{FDM}, only a single point surpasses decoherence accuracy, with a matrix of dimension $499 \times 499$, far beyond the range of Fig.~\ref{fig:LCmetrics}a. In Fig.~\ref{fig:LCmetrics}b, we see that even the smallest grid sizes do not reach decoherence accuracy, and saturate at accuracy levels far larger than any of the DVRs. As we highlight in Appendix~\ref{FDM}, this saturation also occurs at far larger matrix sizes, and while some improvement in $P$ can be found by utilizing higher-order approximations of the finite difference method, this does not yield dramatic improvement in matrix size and reduces the most attractive quality of this approach --- its sparseness. This behavior is common to all the examples we consider, hence, it is excluded in our results going forward.

\subsection{Fluxonium}
\label{Fluxonium}

Of particular interest is the fluxonium~\cite{fluxonium,fluxoniumEX, companion, fluxcoherence2}, shown in Fig.~\ref{fig:examples}b, which has no exact analytical solution and whose standard numerical approach is known to be less than ideal~\cite{fluxonium}. We can model the fluxonium by replacing the capacitance in the LC oscillator with a superconducting tunneling element, the Josephson junction, which has an associated Josephson energy $E_J$ and junction capacitance $C_J$, and demanding that the inductance be large, such that $E_J \gg E_L$. This results in the Hamiltonian
\begin{equation}
\label{eq:FLham}
\hat{H} = 4 E_C \hat{N}^2 + \frac{E_L}{2} \hat{\theta}^2 - E_J \cos (\hat{\theta} + 2 \pi \mathcal{A}) \enspace ,
\end{equation}
where analogous to the LC oscillator, we have the charging energy $E_C = \frac{e^2}{2 C_J}$ and the inductive energy $E_L = \frac{1}{L} \left(\frac{\Phi_0}{2 \pi}\right)^2$. The fluxonium is also threaded by an external flux $\Phi^{\text{ext}}$, captured by the parameter $\mathcal{A} = \frac{\Phi^{\text{ext}}}{\Phi_0}$, with $e$ the elementary electron charge and $\Phi_0$ the magnetic flux quantum.

The standard numerical approach~\cite{fluxonium, Groszkowski2021scqubitspython} consists of representing the fluxonium's Hamiltonian using a harmonic oscillator basis that diagonalizes Eq.~(\ref{eq:FLham})'s first two terms. This is described in more detail in Appendix~\ref{FLextra}. However, this method is known to require larger state spaces for convergence, as the harmonic oscillator basis is not a good approximation for the fluxonium's eigenstates. Thus, we are motivated to consider alternative approaches for a numerical representation of this system. Like the LC oscillator, the fluxonium has a bounded, non-periodic potential, and as a result, there are no requirements on a variable's periodicity to restrict the usage of DVRs. Therefore, we consider all four of the DVRs presented in Section~\ref{S2} for comparison with each other and to the standard numerical method, exploring their features and performance in the context of level convergence, basis state contributions, and phase shifts.

\subsubsection{Level convergence}

We assess the energy level convergence properties of each DVR by examining the absolute value of the energy difference, $\Delta_n$ [Eq.~(\ref{eq:ED})], as a function of matrix size, i.e., the number of DVR basis states included in a numerical representation. However, for the fluxonium, no exact solution exists to serve as the reference energy $E^{\rm ref}_n$ in Eq.~(\ref{eq:ED}). Instead, in this example, we take $E^{\rm ref}_n$ to be that provided by the standard numerical approach using a sizable number of basis states. Thus, we represent the fluxonium using the standard harmonic oscillator basis [Eqs.(\ref{eq:HObasis1})-(\ref{eq:HObasis2})] with the length scale $\theta_0 = (8 E_C/E_L)^{1/4}$ and 1001 basis states, assuring that the lowest levels of the system are well-converged to serve as a reference point for the DVRs as well as the harmonic oscillator basis with smaller matrix sizes and alternative length scales.

Analogously to the LC oscillator, we consider various grid sizes in all four DVRs, employing the use of the decoherence-accurate matrix size $R$ and saturation precision $P$ to assess each DVR's relative performance. A key distinction in this example is that for the charge DVRs, we consider only those grid sizes for which $1/\Delta N$ is an integer, yielding a trivial form for the contribution from the cosine operator, as discussed in Section~\ref{OpRep}. Appendix~\ref{FLextra} includes visuals of $|\Delta_0|$ as a function of matrix size for a sampling of grid sizes, from which we extract $R$ and $P$, as well as additional details of the numerical analysis. We note that while there exists some functional variation in $|\Delta_0|$ upon comparison with the LC oscillator case (which we attribute to the oscillatory contribution to the potential), the overall behavior of these curves with respect to grid size, as described in Section~\ref{LCoscillator}, is the same.

\begin{figure}
\centering
\includegraphics[width= 0.48\textwidth]{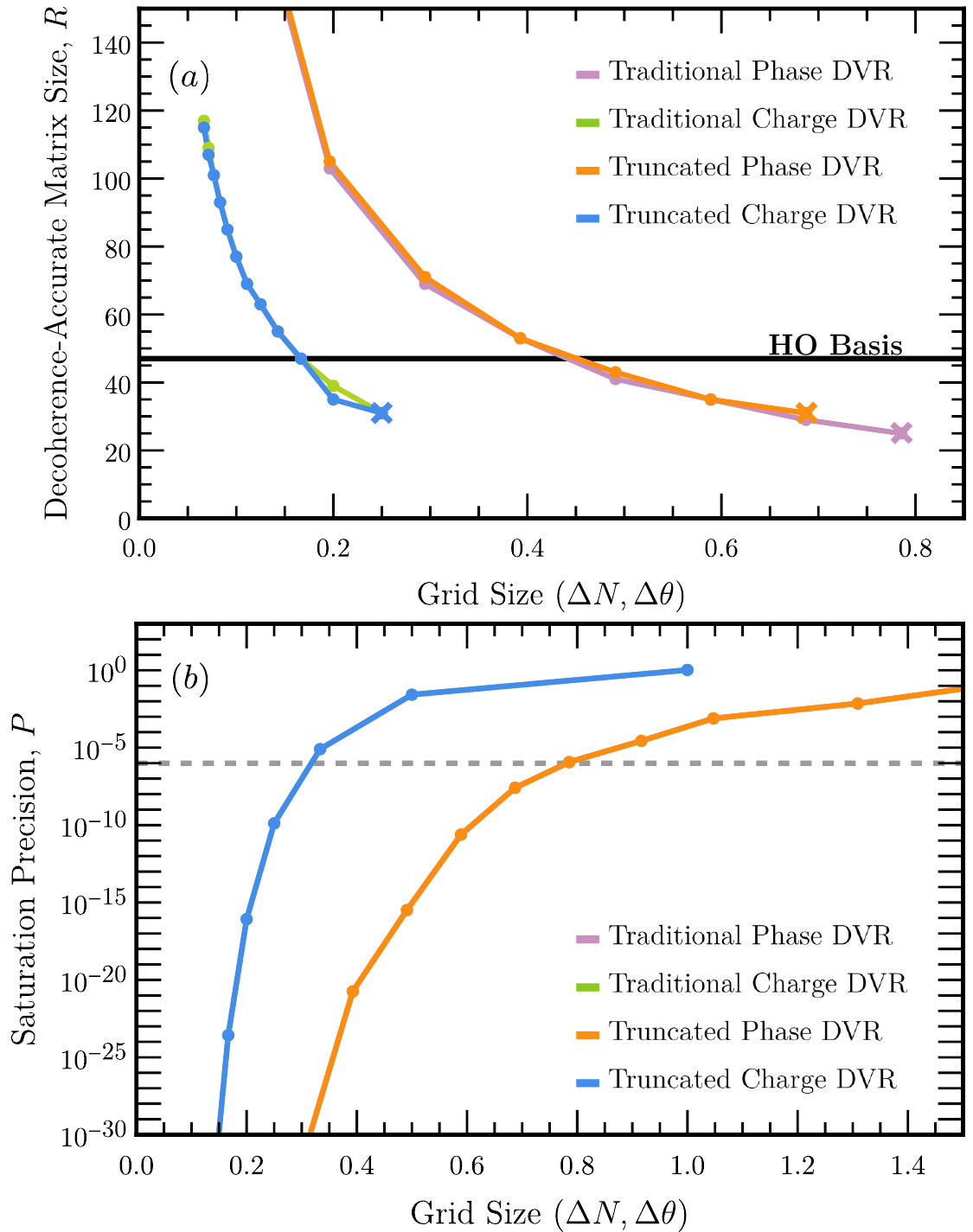}
\caption{(a) The decoherence-accurate matrix size $R$ as a function of grid size for the fluxonium ground state, for all four of the DVRs considered and colored according to the included legend. The matrix size at which the standard numerical approach, the harmonic oscillator basis, reaches decoherence accuracy is indicated by a horizontal solid black line. The grid size cutoff for $R$, i.e., the largest grid size that surpasses decoherence accuracy, is indicated with an `X' data point. (b) The saturation precision $P$ as a function of grid size for the fluxonium ground state, for all four of the DVRs considered. Decoherence accuracy is indicated with a horizontal gray dashed line. Parameters for these plots are $E_C/h = 2.5$ GHz, $E_L/h = 0.5$ GHz, $E_J/h = 10$ GHz, and $\mathcal{A} = 1/2$.}
\label{fig:FLmetrics}
\end{figure}

We plot the decoherence-acccurate matrix size $R$ and the saturation precision $P$ as a function of a DVR's grid size in Figs.~\ref{fig:FLmetrics}a and \ref{fig:FLmetrics}b, respectively, specifically for the ground state of the fluxonium. Overall, we find behavior that is analogous to that found in the LC oscillator. Fig.~\ref{fig:FLmetrics}a demonstrates an overall increase in $R$ as the grid size decreases, for all four DVRs. As the grid size increases, $R$ decreases, ultimately reaching a cutoff beyond which decoherence accuracy is never reached. In Fig.~\ref{fig:FLmetrics}b, we see again how larger grid sizes tend towards saturation at worse accuracy, while smaller grid sizes lead to rapid improvements, saturating below decoherence accuracy in numerous cases.

While $P$ is nearly identical across the traditional and truncated DVRs, as anticipated, we see slightly more variation in $R$. We also find smaller grid size cutoffs for $R$ in the fluxonium, where grid sizes larger than $\Delta N = 0.25$ and $\Delta \theta = \pi/4, 7\pi/32$ do not achieve decoherence accuracy. Notably, the phase DVRs have different grid size cutoffs in $R$. However, in the traditional phase DVR, the largest grid size to achieve decoherence accuracy is one that, according to Fig.~\ref{fig:FLmetrics}b, saturates barely \textit{above} decoherence accuracy. This discrepancy is a manifestation of the non-variational nature of DVRs, where in the process of converging below the reference energy, high accuracy (below decoherence accuracy) is achieved before ultimately converging at a worse accuracy level, further from the reference energy. Thus, these differing cutoffs are a consequence of both the non-variational nature of DVRs in general as well as the close proximity of this particular grid size's saturation precision to decoherence accuracy. In total, these variations between the truncated and traditional DVRs are still quite small, indicating that the approximation made in the traditional case does not introduce very large error in this system.

Included in Fig.~\ref{fig:FLmetrics}a is a horizontal black line indicating the matrix size at which the standard numerical approach, the harmonic oscillator basis, first surpasses decoherence accuracy. Importantly, there are numerous grid sizes, particularly in the phase DVRs, which achieve this level of accuracy at noticeably smaller matrix sizes. We also note that the harmonic oscillator basis does not saturate --- hence no point of comparison in Fig.~\ref{fig:FLmetrics}b --- but continues to increase in accuracy with increased matrix size. As we highlight further in Appendix~\ref{FLextra} using Fig.~\ref{fig:FLcurves}, level convergence is achieved in all four DVRs, with various grid sizes, at consistently higher accuracy and smaller matrix sizes than the harmonic oscillator basis. This suggests that both the phase and charge number DVR's basis states are better approximations to the fluxonium's eigenstates than the harmonic oscillator basis, potentially providing an alternative numerical approach that is a more efficient and accurate representation for the fluxonium than current standard methods.

This improvement can be made more tangible by considering the impact of smaller matrix sizes on the computational cost of matrix diagonalization and the maximum number of qubits capable of being simulated by a given machine. Consider a system with $\eta$ degrees of freedom, i.e., qubits, where each qubit is described by a Hamiltonian matrix with dimension $R$, the smallest matrix size to achieve decoherence accuracy. The diagonalization of the entire system, neglecting sparseness from the tensor product structure, has a computational cost that scales as $\mathcal{O}(R^{3\eta})$, as matrix diagonalization is an operation that scales as $\mathcal{O}(\mathcal{N}^3)$ with $\mathcal{N} = R^\eta$, the matrix dimension of the entire system. With the decoherence-accurate matrix sizes indicated in Fig.~\ref{fig:FLmetrics}a, $R_{\rm HO} = 47$ and  $R_{\rm DVR} = 31$, we find the speed up provided by DVRs is given by the factor $(R_{\rm DVR}/R_{\rm HO})^{3\eta} \approx (2/3)^{3\eta}$, which only increases with more degrees of freedom. Alternatively, we can consider the largest number of qubits a given machine can represent in its memory, given by $\eta^{\rm max} = \log \mathcal{N}^{\rm max} / \log R$, where $\mathcal{N}^{\rm max}$ corresponds to the largest matrix dimension capable of being diagonalized. While the ratio $\eta^{\rm max}_{\rm DVR}/\eta^{\rm max}_{\rm HO}$ scales logarithmically and $R_{\rm HO}$ and $R_{\rm DVR}$ are of the same order, the improvements offered by DVRs in this instance still provide about 12\% more qubits.

\begin{figure}
\centering
\includegraphics[width= 0.48\textwidth]{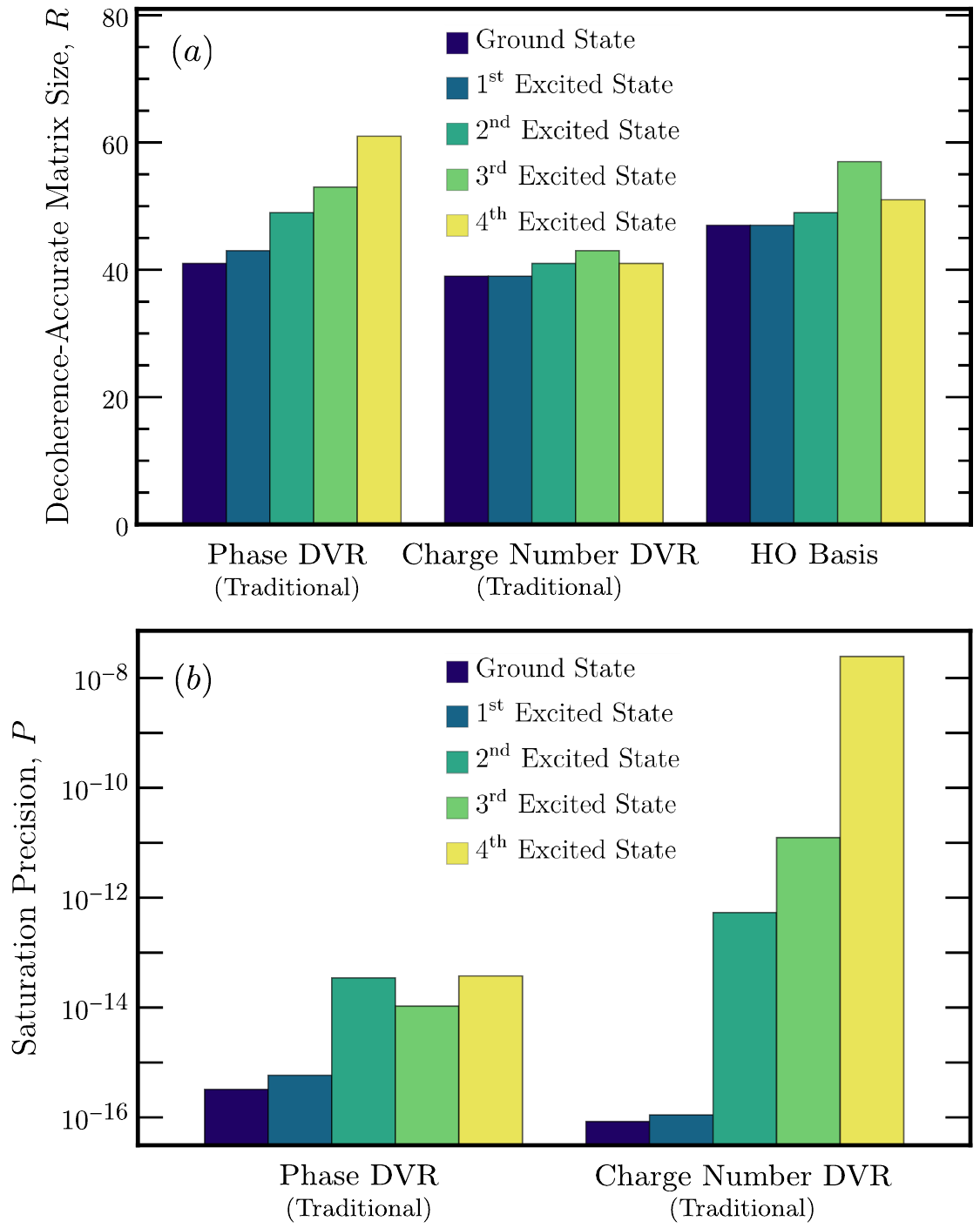}
\caption{(a) The decoherence-accurate matrix size $R$ for the first five energy eigenstates of the fluxonium in the traditional phase DVR ($\Delta \theta = 5 \pi /32$), the traditional charge number DVR ($\Delta N = 1/5$), and the harmonic oscillator basis. (b) The saturation precision $P$ for the first five energy eigenstates of the fluxonium in the traditional phase DVR ($\Delta \theta = 5 \pi /32$) and the traditional charge number DVR ($\Delta N = 1/5$). Parameters for these plots are $E_C/h = 2.5$ GHz, $E_L/h = 0.5$ GHz, $E_J/h = 10$ GHz, and $\mathcal{A} = 1/2$.}
\label{fig:FLHighEnergy}
\end{figure}

Thus far, we have only considered the ground state of the LC oscillator and the fluxonium in our assessment of a DVR's performance. However, in the fluxonium, energy levels beyond the ground state are particularly relevant. Thus, in Fig.~\ref{fig:FLHighEnergy}, we examine the decoherence-accurate matrix size $R$ and the saturation precision $P$ for the first five energy eigenstates of the fluxonium. In Fig.~\ref{fig:FLHighEnergy}a we compare $R$ across the first five levels in the traditional phase and charge number DVRs as well as the harmonic oscillator basis. In particular, we choose the grid sizes $\Delta N = 1/5$ and $\Delta \theta = 5 \pi /32$ for comparison, as they saturate to similar levels of accuracy, well below decoherence accuracy, and are the smallest grid sizes that outperform the harmonic oscillator basis in the ground state with regard to $R$. Given the similarity between the traditional and truncated DVRs of each type, we opt to show only the traditional case in Fig.~\ref{fig:FLHighEnergy}.

For both $R$ and $P$, Fig.~\ref{fig:FLHighEnergy} indicates a degradation in performance whose severity varies depending on the basis used to represent the fluxonium. With respect to $R$, Fig.~\ref{fig:FLHighEnergy}a shows that the phase DVR exhibits the largest drop in performance, while the charge number DVR is relatively stable with respect to increases in energy level. The harmonic oscillator basis similarly degrades in performance for higher energy levels, but not to the extent of the phase DVR. Notably, the charge number DVR outperforms the harmonic oscillator basis, with smaller values of $R$ at all energy levels. Examining the saturation precision $P$ in Fig.~\ref{fig:FLHighEnergy}b, we instead find that while the phase DVR's saturation precision does worsen at higher energy levels, it is relatively stable when compared to the charge number DVR, which demonstrates a far more severe degradation in performance. Ultimately, if considering the use of either DVR to numerically represent the fluxonium, choices in grid size and matrix size must be made to mitigate the influence of this degradation in performance. However, even with this drop in performance at higher energy, the charge number and phase DVRs can outperform the harmonic oscillator basis.

\subsubsection{Basis state contributions}

In addition to comparing the level convergence properties of each DVR against that of the harmonic oscillator basis, we also examine and compare the energy eigenstate decompositions associated with each approach, aiming to characterize the size of each basis state's contribution to an energy eigenstate when represented in a given basis. We accomplish this by considering the magnitude of the overlap of the $i^{\rm th}$ energy eigenstate of the fluxonium $\ket{\Psi_i}$ and the $\alpha$-indexed state of a particular basis $\ket{\psi_{\alpha}}$, i.e., $|\bra{\Psi_i}\ket{\psi_{\alpha}}|^2$. Here, we compare eigenstate decompositions in terms of the traditional phase and charge number DVRs, as well as the harmonic oscillator basis. Once again, given the similarity we have demonstrated between the traditional and truncated DVRs, we opt to focus on only the traditional case here.

To examine the eigenstate contributions from the traditional phase and charge number DVRs, we restrict ourselves to a single instance of each, i.e., one grid size and matrix size per DVR. For each grid size, we choose $\Delta \theta = 5 \pi /32$ for the phase DVR and $\Delta N = 1/5$ for the charge number DVR. At these grid sizes, a state space of 81 basis states is sufficient for both DVRs to have converged to saturation precision in the ground state, which is of similar accuracy in both cases. This can be confirmed in Fig.~\ref{fig:FLcurves} of Appendix~\ref{FLextra}. For appropriate comparison, we also consider the harmonic oscillator basis with a state space of equivalent size.

\begin{figure*}
\centering
\includegraphics[width= 0.96\textwidth]{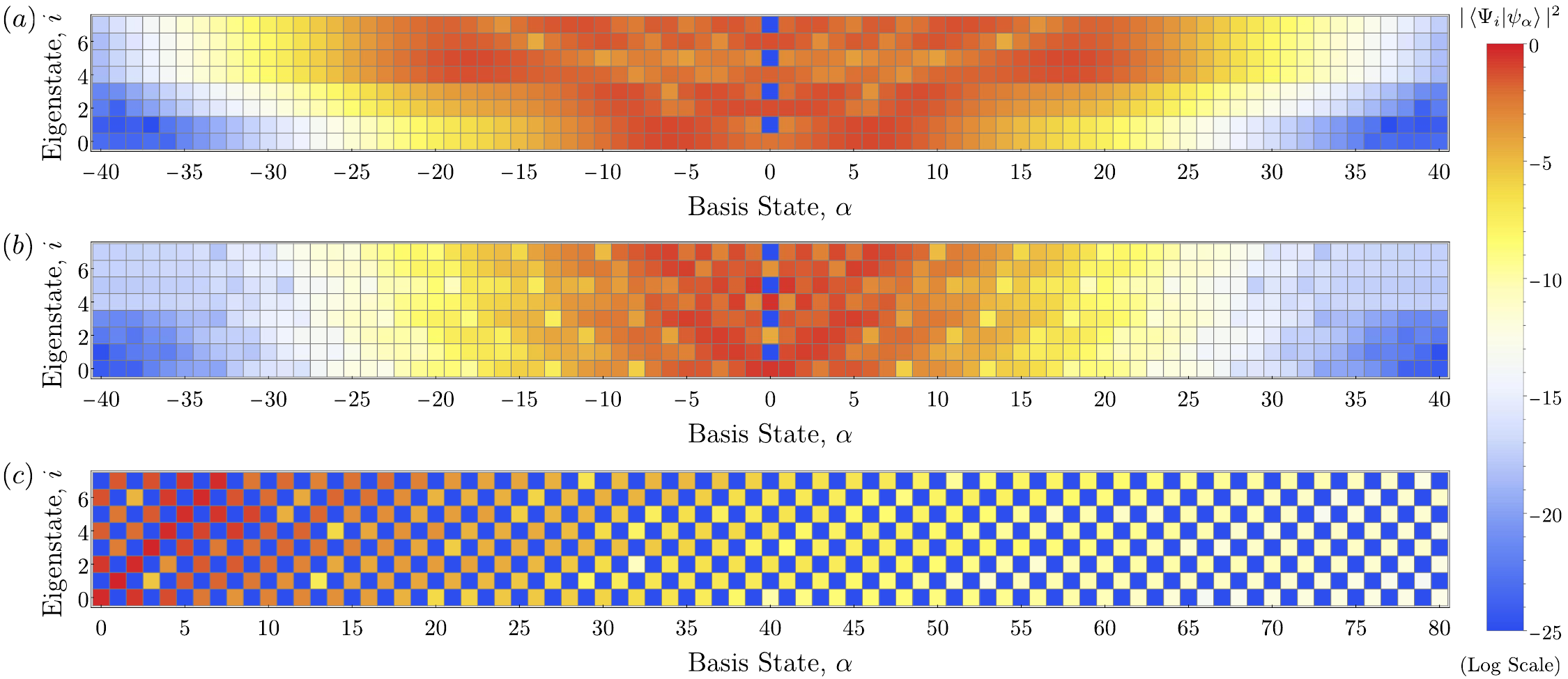}
\caption{Eigenstate decompositions showing $|\bra{\Psi_i}\ket{\psi_{\alpha}}|^2$ for the first eight energy eigenstates of the fluxonium across a state space of 81 basis states for (a) the traditional phase DVR ($\Delta \theta = 5 \pi /32$), (b) the traditional charge number DVR ($\Delta N = 1/5$), and (c) the harmonic oscillator basis. Values are shown on a log scale, colored according to their power, as indicated by the legend. We enforce a cutoff at $10^{-25}$, hence grid spaces with values smaller than this cutoff are automatically colored blue. Parameters for these plots are $E_C/h = 2.5$ GHz, $E_L/h = 0.5$ GHz, $E_J/h = 10$ GHz, and $\mathcal{A} = 1/2$.}
\label{fig:BasisCons}
\end{figure*}

In Fig.~\ref{fig:BasisCons}, we show $|\bra{\Psi_i}\ket{\psi_{\alpha}}|^2$ for the first eight energy eigenstates of the fluxonium across a state space of 81 basis states of the traditional phase DVR, the traditional charge number DVR, and the harmonic oscillator basis. In all cases, these higher energy levels have reached decoherence accuracy and are considered sufficiently well-approximated. It is immediately clear from Fig.~\ref{fig:BasisCons}c that the contributions to an eigenstate represented in the harmonic oscillator basis are markedly different from those represented with either the phase or charge number DVRs, shown in Fig.~\ref{fig:BasisCons}a or Fig.~\ref{fig:BasisCons}b, respectively. With the harmonic oscillator basis, only states of common parity contribute to a given eigenstate, permitting one to effectively neglect half of the defined state space. This is not present with either DVR, however, we do find an analogous effect specifically with regard to the contribution from the central basis state of each DVR --- its contribution (or lack thereof) alternates with increasing energy.

Notably, Fig.~\ref{fig:BasisCons}b indicates that the traditional charge number DVR is a particularly compact representation. The main contributions to an eigenstate come from the central 10 to 20 states, which is consistently true even at higher energy. The drop in magnitude beyond this range is dramatic, with minimal contributions coming from the edge states. The phase DVR exhibits a similar drop in magnitude away from the central contributing states, however, those states vary considerably more with increased energy. While low-lying levels are relatively compact, with comparable numbers of basis states contributing largely, as in the charge number DVR, higher energy levels demonstrate considerable broadening in this regard. The harmonic oscillator basis appears to fall somewhere in between --- its main contributing states are certainly focused about the lowest harmonic oscillator states, which increase slightly with increased energy. However, the drop in magnitude away from these main contributing states is considerably slower than in either DVR. This may point to the harmonic oscillator's worse performance with regard to accuracy and matrix size, indicating far more states are required to reach equivalent levels of accuracy achieved with DVRs.

\subsubsection{Phase shifts}

A specific application for which the traditional and truncated phase DVRs are particularly well suited is the implementation of phase shifts in a system's wavefunction. The construction of these DVRs results in the discretization of the phase variable, enabling a convenient mathematical definition for such a phase-shift operation. As we now show, with a straightforward choice for the size of the phase shift, we need only shift a state's expansion coefficients relative to the DVR's basis states to generate the phase shift.

Consider a system's eigenstate $\ket{\Psi_n}$ expanded in terms of either phase DVR,
\begin{equation}
\ket{\Psi_n} = \sum_\alpha c_{\alpha n} \ket{\psi_{\alpha,\theta}^{(A,B)}} \enspace ,
\end{equation}
where $c_{\alpha n}$ is the expansion coefficient corresponding to the overlap between the eigenstate and each basis state, $\bra{\psi_{\alpha,\theta}^{(A,B)}}\ket{\Psi_n}$. In a continuous phase representation, where $f(\theta) = \bra{\theta} \ket{f}$, this expansion becomes
\begin{equation}
\Psi_n(\theta) = \sum_\alpha c_{\alpha n} \psi_{\alpha,\theta}^{(A,B)} (\theta) \enspace .
\end{equation}
Implementing a phase shift $\pm \phi$, like with any other basis, is then accomplished by taking $\theta \rightarrow \theta \pm \phi$, yielding
\begin{equation}
\label{eq:PSfunc}
\Psi^\pm_n (\theta) \equiv \Psi_n(\theta \pm \phi) = \sum_\alpha c_{\alpha n} \psi_{\alpha,\theta}^{(A,B)} (\theta \pm \phi) \enspace  .
\end{equation}
This procedure is greatly simplified, however, by noting that the $\alpha$-indexed basis states of a phase DVR are only a phase shift of the grid size $\Delta \theta$ away from each other. Therefore, if the phase shift $\pm \phi$ can be written as an integer multiple of the DVR's grid size, i.e., $\pm \phi = \pm \beta \Delta \theta$ with $\beta$ a positive integer, we can express the phase-shifted basis state in Eq.~(\ref{eq:PSfunc}) as just another indexed basis state, 
\begin{equation}
\psi_{\alpha,\theta}^{(A,B)} (\theta \pm \phi) = \psi_{\kappa,\theta}^{(A,B)} (\theta) \enspace ,
\end{equation}
where $\kappa = \alpha \mp \beta$. The particular direction of the shift relative to $\alpha$ is readily confirmed by examining the forms of $\psi_{\alpha,\theta}^{(A,B)} (\theta \pm \phi)$ [Eqs.~(\ref{eq:tradsincTHF}) and (\ref{eq:truncsincTHF})], where we can re-write the quantity $\pm \phi - \theta_\alpha = -(\alpha \mp \beta) \Delta \theta = -\kappa \Delta \theta = -\theta_\kappa$.

The phase-shifted eigenstate $\Psi^\pm_n (\theta)$ can then be written conveniently in Dirac notation as
\begin{equation}
\ket{\Psi_n^\pm} = \sum_\alpha c_{\alpha n} \ket{\psi_{\kappa,\theta}^{(A,B)}} \enspace ,
\end{equation}
where we see that these two phase DVRs accomplish a phase shift $\pm \phi = \pm \beta \Delta \theta$ via a corresponding shift in the DVR's basis states relative to their expansion coefficients. Further, this straightforward implementation of a phase shift in the traditional and truncated phase DVRs enables the identification of a phase-shift operator $\hat{\mathcal{O}}_\pm^{(A,B)}$,
\begin{equation}
\begin{split}
\ket{\Psi_n^\pm} &= \sum_\alpha c_{\alpha n} \ket{\psi_{\kappa,\theta}^{(A,B)}} \\
&= \sum_\alpha  \ket{\psi_{\alpha \mp \beta,\theta}^{(A,B)}} \bra{\psi_{\alpha,\theta}^{(A,B)}}\ket{\Psi_n} \\
&= \hat{\mathcal{O}}_\pm^{(A,B)} \ket{\Psi_n} \enspace ,
\end{split}
\end{equation}
which we can re-express as iterations of a single matrix,
\begin{equation}
\label{eq:PSoperator}
\begin{split}
\hat{\mathcal{O}}_\pm^{(A,B)} &= \sum_\alpha \ket{\psi_{\alpha \mp \beta,\theta}^{(A,B)}} \bra{\psi_{\alpha,\theta}^{(A,B)}} \\
&= \left( \sum_\alpha \ket{\psi_{\alpha \mp1,\theta}^{(A,B)}} \bra{\psi_{\alpha,\theta}^{(A,B)}} \right)^\beta \enspace .
\end{split}
\end{equation}
The matrix inside the parenthesis represents a raising or lowering operator that connects the basis states of a phase DVR --- essentially carrying out the $\Delta \theta$ phase shift that separates them. By multiplying this matrix $\beta$ times, we generate the desired phase shift $\pm \phi = \pm \beta \Delta \theta$, as expected. However, we must be mindful of the specific characteristics of each phase DVR that influence the form of this matrix.

Recall that the traditional sinc DVR is a basis of infinite size, where $\alpha = -\infty, ..., \infty$. Thus, for an eigenstate expansion written in terms of the traditional phase DVR, the phase-shift operator $\hat{\mathcal{O}}_\pm^{A}$ is an infinite-dimensional matrix. In theory, $\beta$ can be any value, with $\hat{\mathcal{O}}_\pm^{A}$ capable of implementing phase shifts of arbitrary size. In practice, however, this matrix is truncated at some finite size such that the function of interest and its phase shift are well-approximated, i.e., contributions from the basis states at the numerically-imposed boundary are near zero, and $\beta$ is much smaller than the total size of the state space. We note that upon implementing the phase-shift operator in this truncated state space, the boundary terms appearing in the iterated matrix in Eq.~(\ref{eq:PSoperator}) are set to zero, and a $\beta$-number of coefficients are shifted outside of the state space and replaced with zero. However, so long as the state space is large enough, this introduces no appreciable errors, which can be numerically confirmed by checking the state's norm.

For an energy eigenstate expansion written in terms of the truncated phase DVR, the phase-shift operator $\hat{\mathcal{O}}_\pm^{B}$ is instead a finite-sized matrix with dimension $2M+1$, as this basis is finite-sized by definition with $\alpha = -M,...,M$. Importantly, the basis states of the truncated phase DVR are $2 \theta_{\rm max}$-periodic, corresponding to a phase shift with $\beta = 2M+1$. Therefore, the boundary terms appearing in the iterated matrix in Eq.~(\ref{eq:PSoperator}) map back inside the interval $[-M,M]$. While $\beta$ can take on any value, the resulting phase shift, characterized by $\kappa$, always maps back to a state indexed inside $[-M,M]$, resulting in a net phase shift no larger than $2 \theta_{\rm max}$. As expected, $\beta = 2M+1$ yields no net phase shift and the phase-shift operator is the identity, $\hat{\mathcal{O}}_\pm^{B} = \mathbb{1}$. As in the traditional case, in practice we choose the size of the state space, determined by $M$, to be large enough to well-approximate a function of interest as well as its phase shift, which is particularly relevant when choosing to represent and generate phase shifts with a non-periodic function using this basis.

\begin{figure}[!t]
\centering
\includegraphics[width= 0.48\textwidth]{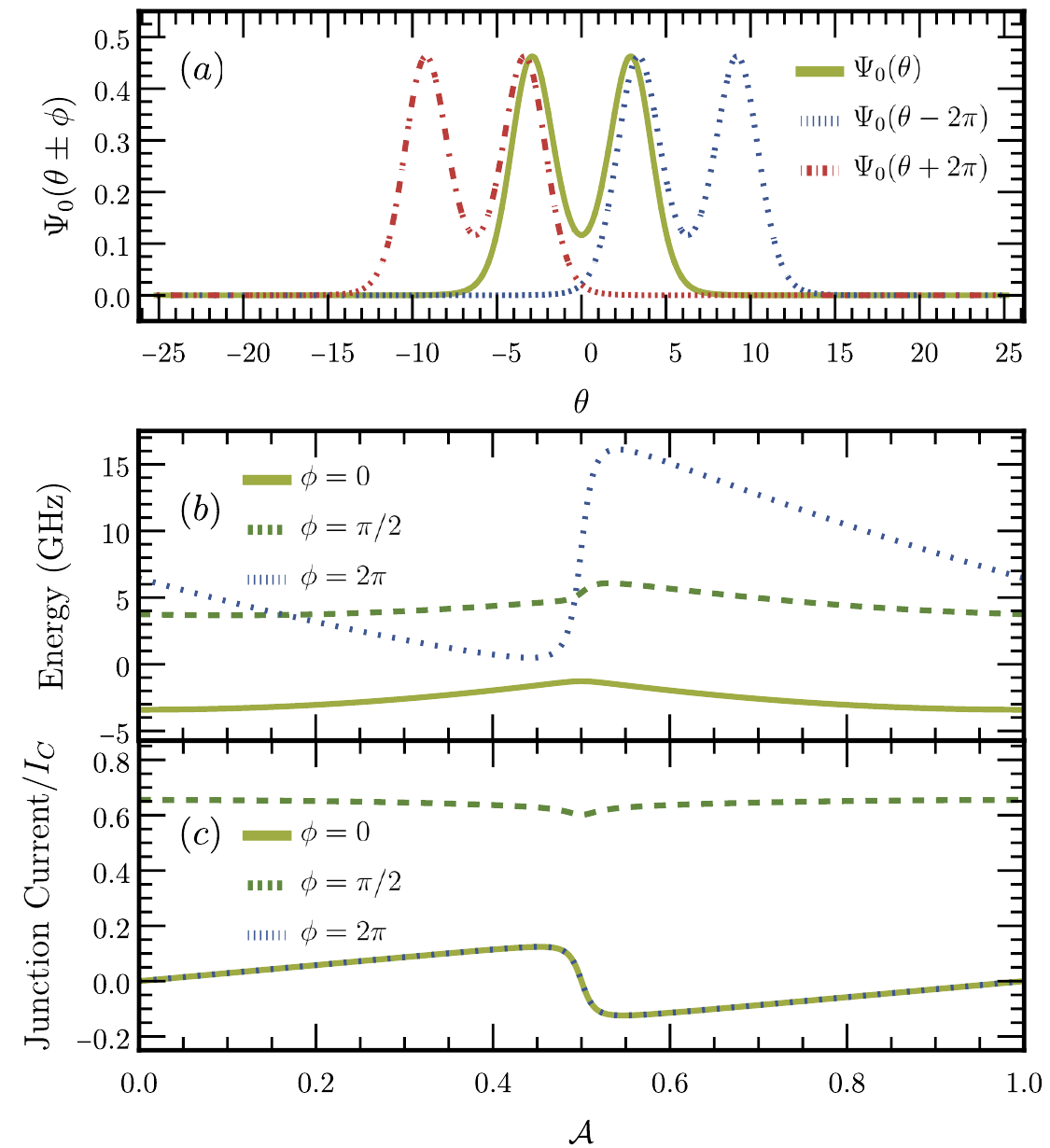}
\caption{(a) Visualization of the phase shift operation in the ground state of the fluxonium with a flux penetration corresponding to $\mathcal{A} = 1/2$, specifically comparing the original wavefunction to the forward and backward phase-shifted wavefunctions with a phase shift of $2 \pi$. In (b) and (c), we specifically focus on the forward-shifted state $\Psi_0(\theta - \phi)$ with phase shifts of $\phi = 0, \pi/2, 2 \pi$. In (b), we plot the average energy of the state $\bra{\Psi_0^-} \hat{H} \ket{\Psi_0^-}$ as a function of the external flux while in (c), we plot the supercurrent through the Josephson junction $\bra{\Psi_0^-} \hat{I}/I_C \ket{\Psi_0^-}$ as a function of the external flux, where $\hat{I} = I_C \sin (\hat{\theta} + 2 \pi \mathcal{A})$, with $I_C$ the junction's critical current. Parameters for these plots are $E_C/h = 2.5$ GHz, $E_L/h = 0.5$ GHz, and $E_J/h = 10$ GHz. All phase-shifted functions are generated via the methods outlined in the main text.}
\label{fig:PS}
\end{figure}

The identification of the matrix in Eq.~(\ref{eq:PSoperator}) as a phase shift operator yields a convenient method for manipulating state vectors numerically to generate a phase shift, without requiring an explicit functional form. This allows for a straightforward examination of the impact of such phase shifts on a system. As an example, we include Fig.~\ref{fig:PS}, which shows the effects of phase shifts in the fluxonium, specifically for the ground state wavefunction, its energy, and its associated currents. We generate phase shifts in the fluxonium ground state using the traditional phase DVR, with a grid size $\Delta \theta = \pi/8$ and a matrix size corresponding to 101 states. This is sufficiently large enough to accurately represent the ground state (see Fig.~\ref{fig:FLcurves}a) as well as any associated phase shift we implement here, which can be confirmed numerically via the state norm.

Fig.~\ref{fig:PS}a shows the ground state wavefunction $\Psi_0(\theta)$ of the fluxonium for an external flux threading corresponding to $\mathcal{A} = 1/2$, as well as its phase-shifted counterparts $\Psi_0(\theta \pm \phi)$ for $\phi = 2 \pi$, generated via the phase shift operator in Eq.~(\ref{eq:PSoperator}). Phase-shifted states such as those shown in Fig.~\ref{fig:PS}a can then be utilized to consider the effect of such a phase shift on the energy of the ground state. In Fig.~\ref{fig:PS}b, we plot the average energy in each of these phase-shifted states as a function of the flux penetration, evaluating the energy by taking the expectation value of the fluxonium Hamilonian in Eq.~(\ref{eq:FLham}) with the phase-shifted states, i.e., $\bra{\Psi_0^-} \hat{H} \ket{\Psi_0^-}$, for $\phi = 0,\pi/2,2 \pi$. We can similarly examine the effect of the phase shift on the current through the Josephson junction, shown in Fig.~\ref{fig:PS}c, by instead taking the expectation value of the operator corresponding to the supercurrent through the junction, $\hat{I} = I_C \sin (\hat{\theta} + 2 \pi \mathcal{A})$, where $I_C$ represents the junction's critical current and the operator $\hat{I}$ is constructed analogously to those presented in Section~\ref{OpRep}. In both instances, we observe increases in energy and current due to these manufactured phase shifts, which vary in size as a function of the flux penetration as well as the size of the phase shift. Importantly, we demonstrate the applicability of phase DVRs for generating and modeling the impacts of phase shifts in systems like the fluxonium.

\subsection{Transmon}
\label{transmon}

One standard circuit that presents a unique, albeit more restrictive, use-case for DVRs is the transmon/Cooper-pair box~\cite{transmon,devsum,companion,TF,CPB}, shown in Fig.~\ref{fig:examples}c. This system's Hamiltonian is written as
\begin{equation}
\label{eq:TRham}
\hat{H} = 4 E_C (\hat{N} - N_g)^2 - E_J \cos\hat{\theta} \enspace ,
\end{equation}
where the characteristic energy scales are set by the charging energy $E_C = \frac{e^2}{2 (C_J + C_g)}$ and the Josephson energy $E_J$. In the transmon limit ($E_J \gg E_C$), we have the usual transmon qubit, while in the charge limit ($E_J \sim E_C$), we have the well-known Cooper-pair box. Here, the capacitance $C_g$ couples an offset voltage $V_g$ to the circuit, leading to the offset charge $N_g = \frac{C_g V_g}{2 e}$, with $e$ the elementary electron charge. The sensitivity of the system to this offset charge is an identifying feature of these two limits, with the Cooper-pair box notoriously susceptible to offset charge noise.

Notably, this system is $2\pi$-periodic in phase, demanding an integer quantization of charge number. This limits the application of the DVRs we discuss to those which are $2\pi$-periodic in phase with a grid size of $\Delta N = 1$. Therefore, in this example, we can consider only two DVRs: the traditional charge number DVR with $\Delta N = 1$ and $\theta \in [-\pi, \pi]$ and the truncated phase DVR with $\Delta N = 1$ and $\Delta \theta = 2 \pi/(2M+1)$, with $2M+1$ corresponding to the size of the state space. As we discuss in Section~\ref{tradsinc}, the former of these two choices is the usual `charge basis.' However, the latter offers an alternative numerical description in terms of a discretized phase (which depends on the chosen matrix size), allowing for an investigation and comparison of such an approach to the typical method. Here, we examine the level convergence properties and basis state contributions of these two approaches in both energy regimes of the circuit.

\subsubsection{Level convergence}

\begin{figure}
\centering
\includegraphics[width= 0.48\textwidth]{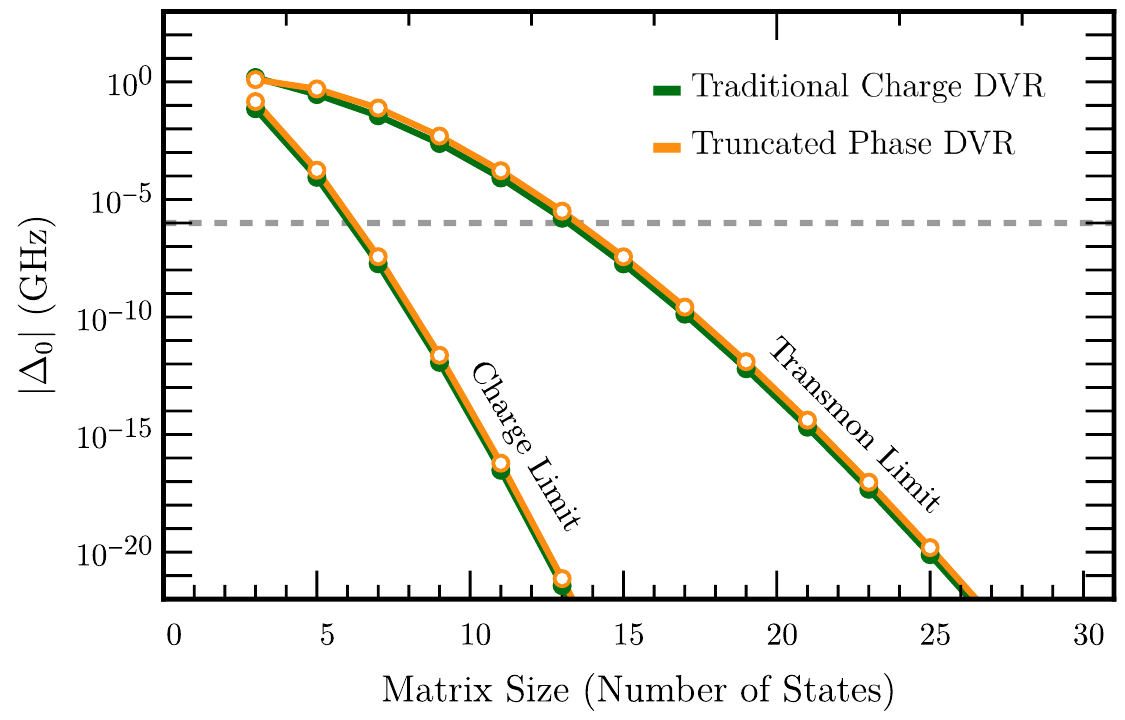}
\caption{The absolute value of the energy difference $|\Delta_0|$ as function of matrix size in the ground state of the transmon in both the transmon limit and the charge limit. Dark green points correspond to the the traditional charge DVR with $\Delta N = 1$, a.k.a. the charge basis. Orange points correspond to the truncated phase DVR with $\Delta N = 1$ and $\Delta \theta$ a function of the matrix size $2M+1$ according to $\Delta \theta = 2 \pi/(2M+1)$. Decoherence accuracy is indicated with a horizontal gray dashed line. Open data points indicate where $\Delta_0 <0$, in order to note and highlight the non-variational nature of DVRs. Parameters for these plots are $E_C/h = 0.2$ GHz, $E_J/h = 10$ GHz, and $N_g = 1/2$ for the transmon limit and $E_C/h = 5$ GHz, $E_J/h = 5$ GHz, and $N_g = 1/2$ for the charge limit.}
\label{fig:TMlvls}
\end{figure}

While the Hamiltonian in Eq.~(\ref{eq:TRham}) is exactly solvable in the continuous phase basis in terms of Mathieu functions and their characteristic values~\cite{transmon,CottetThesis}, these are highly non-trivial functions to work with. As a consequence, standard numerical tools~\cite{Groszkowski2021scqubitspython} typically employ the charge basis with some associated truncation to represent this system. For our analysis, we take $E^{\rm ref}_n$ to be the exact energies~\cite{transmon,CottetThesis} in order to assess the performance of both the charge basis and the truncated phase DVR. This is shown in Fig.~\ref{fig:TMlvls}, where we opt instead to show the explicit level convergence of these two DVRs by examining $|\Delta_0|$ as a function of matrix size for each DVR, as we are limited to only two DVRs with one grid size fixed.

Examining the level convergence of both limits of the transmon in Fig.~\ref{fig:TMlvls}, we immediately observe that the truncated phase DVR, while converging below the true energy, behaves analogously to the charge basis (the traditional charge DVR, $\Delta N = 1$). Additionally, while the truncated phase DVR yields a marginally worse accuracy than the charge basis at each matrix size, this is not large enough to significantly affect performance, as evidenced by their identical decoherence-accurate matrix sizes of 7 states in the charge limit and 15 states in the transmon limit. This highlights the more rapid convergence both exhibit when compared to either of the prior examples considered.  Interestingly, despite the fact that the truncated phase DVR yields a full Hamiltonian matrix and the charge basis results in a sparse, tri-diagonal Hamiltonian matrix (see Section~\ref{OpRep}), the truncated phase DVR still performs on par with the charge basis.

\subsubsection{Basis state contributions}

\begin{figure}[!b]
\centering
\includegraphics[width= 0.48\textwidth]{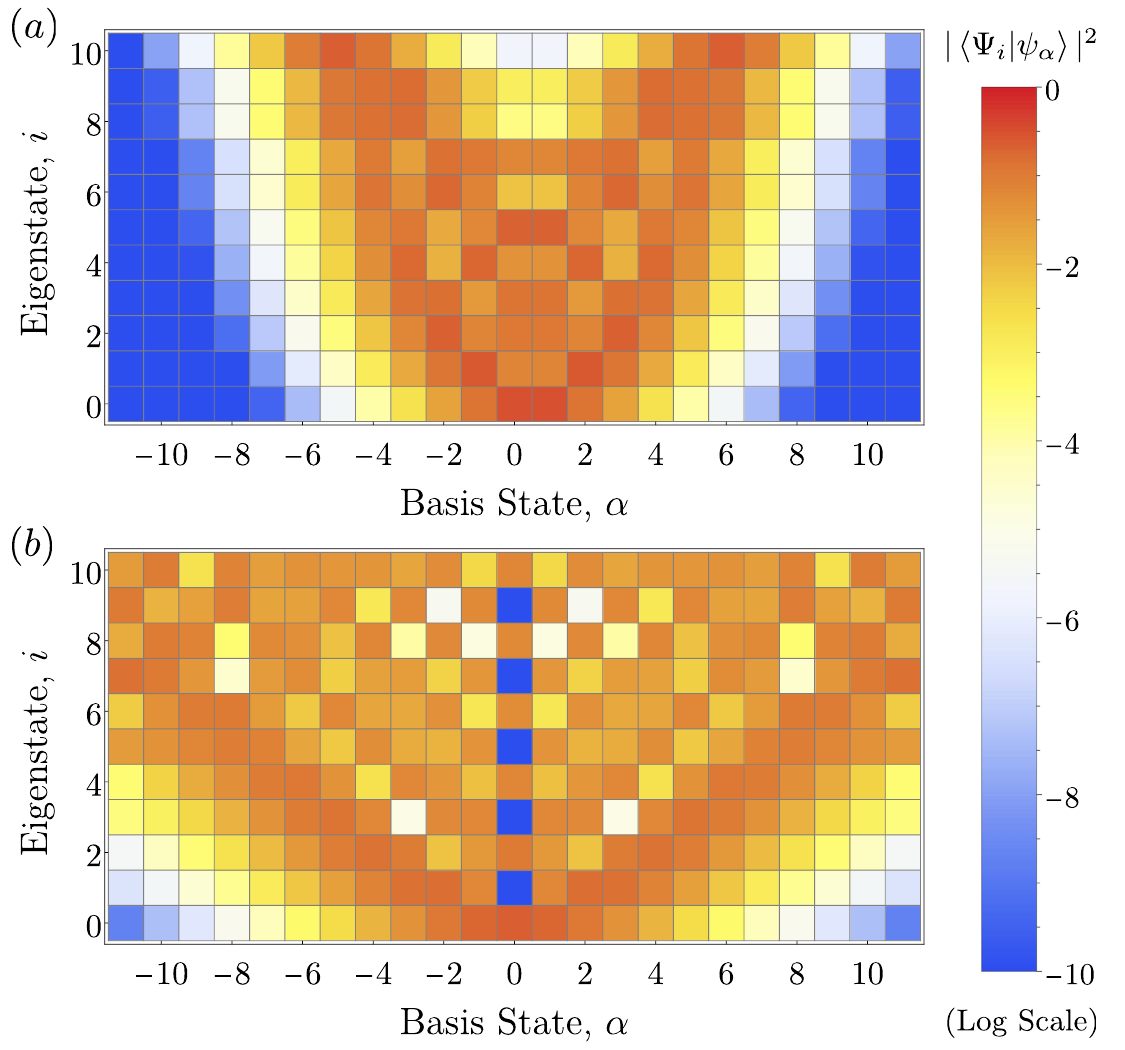}
\caption{Eigenstate decompositions showing $|\bra{\Psi_i}\ket{\psi_{\alpha}}|^2$ for the first eleven energy eigenstates of the transmon in the transmon limit across a state space of 23 basis states in (a) the traditional charge DVR ($\Delta N = 1$) and (b) the truncated phase DVR ($\Delta N = 1$, $\Delta \theta = 2 \pi /23$). Values are shown on a log scale, colored according to their power, as indicated by the legend. We enforce a cutoff at $10^{-10}$, hence grid spaces with values smaller than this cutoff are automatically colored blue. Parameters for these plots are $E_C/h = 0.2$ GHz, $E_J/h = 10$ GHz, and $N_g = 1/2$.}
\label{fig:BasisConTM}
\end{figure}

\begin{figure}[!t]
\centering
\includegraphics[width= 0.48\textwidth]{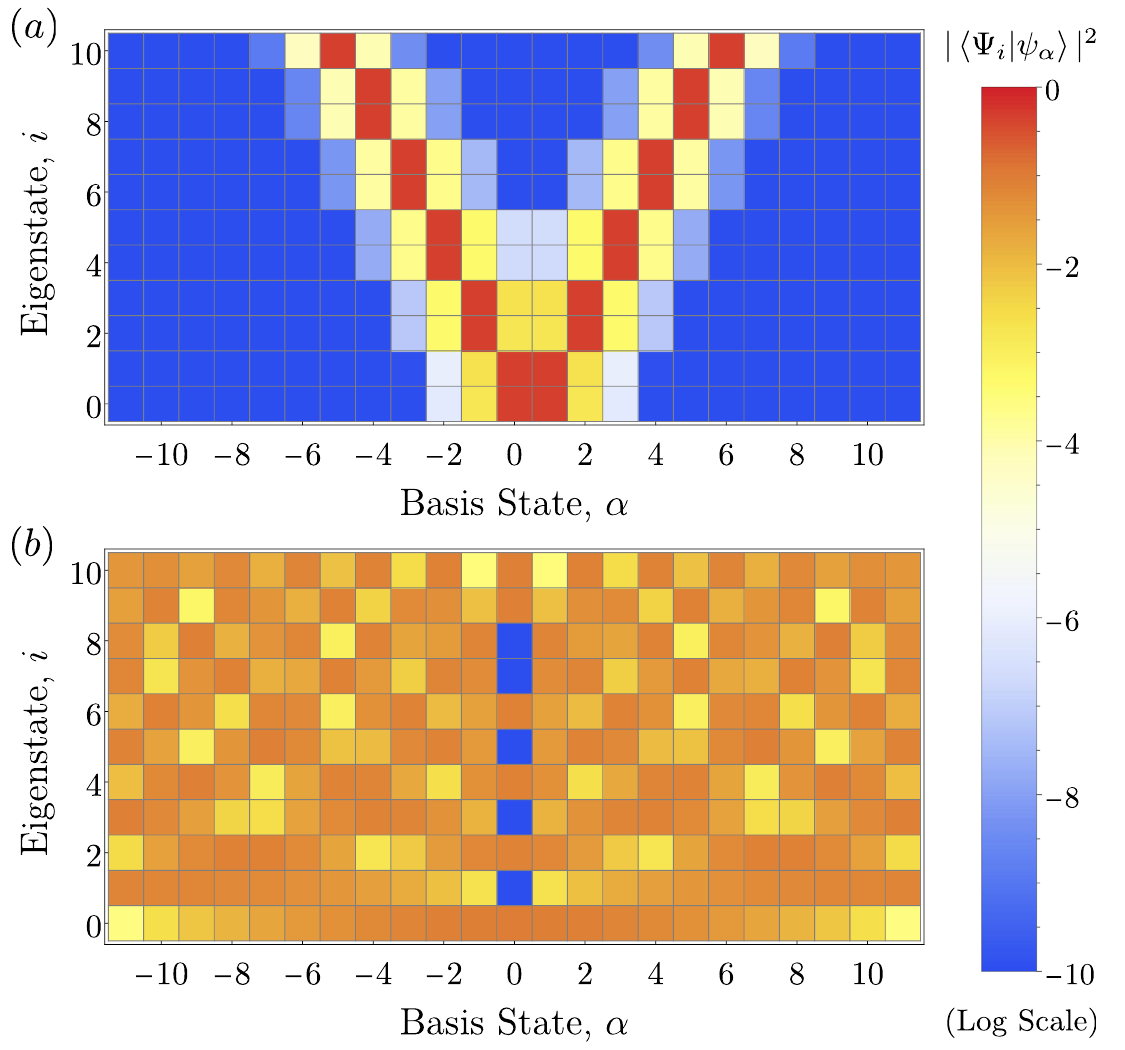}
\caption{Eigenstate decompositions showing $|\bra{\Psi_i}\ket{\psi_{\alpha}}|^2$ for the first eleven energy eigenstates of the transmon in the charge limit across a state space of 23 basis states in (a) the traditional charge DVR ($\Delta N = 1$) and (b) the truncated phase DVR ($\Delta N = 1$, $\Delta \theta = 2 \pi /23$). Values are shown on a log scale, colored according to their power, as indicated by the legend. We enforce a cutoff at $10^{-10}$, hence grid spaces with values smaller than this cutoff are automatically colored blue. Parameters for these plots are $E_C/h = 5$ GHz, $E_J/h = 5$ GHz, and $N_g = 1/2$.}
\label{fig:BasisConCPB}
\end{figure}

We also examine the energy eigenstate decompositions of each approach in either the charge or transmon limit, again evaluating the magnitude of the overlap of the $i^{\rm th}$ energy eigenstate of the transmon $\ket{\Psi_i}$ and the $\alpha$-indexed state of a particular basis $\ket{\psi_{\alpha}}$, i.e., $|\bra{\Psi_i}\ket{\psi_{\alpha}}|^2$. In particular, we consider a state space of 23 basis states, where both DVRs have converged below decoherence accuracy, shown in Fig.~\ref{fig:BasisConTM} and Fig.~\ref{fig:BasisConCPB} for the transmon limit and charge limit, respectively.

From Figs.~\ref{fig:BasisConTM}a and \ref{fig:BasisConCPB}a, it is immediately evident that the charge basis is a far more compact description in either limit. In the charge limit, the charge basis states are very good approximations to the eigenstates of the system, essentially amounting to a superposition of only two charge number states. As $E_J$ grows, moving into the transmon limit, more charge basis states are required to represent an eigenstate. This is a direct reflection of the matrix representation of the Hamiltonian in this basis, where the off-diagonal elements of this tri-diagonal matrix grow in magnitude relative to the diagonal contributions when moving from the charge limit to the transmon limit. On the other hand, using the truncated phase DVR to represent the system requires nearly all of the basis states in a given state space, as shown in Figs.~\ref{fig:BasisConTM}b and \ref{fig:BasisConCPB}b. While in the transmon limit the truncated phase DVR is somewhat localized, in the charge limit this description is not at all compact. As before, this can be understood as a reflection of the Hamiltonian matrix in each limit. The charging contribution to the Hamiltonian is a full matrix in the phase DVR, which grows in relative magnitude upon moving from the transmon limit to the charge limit, overpowering the diagonal contribution from the tunneling term.

\section{Outlook}
\label{outlook}

In this work, we have demonstrated the effectiveness of using sinc DVRs to numerically model superconducting circuits. By considering extensions and generalizations of the usual charge basis, we introduce a broad class of both discrete charge number and discrete phase bases, which can be readily employed in numerous systems that incorporate any combination of capacitors, Josephson junctions, and inductors. Despite their non-variational behavior and lack of sparsity, we find that these DVRs can match, and in many cases outperform, the standard numerical approaches currently used in many software tools today.

We find that these DVRs vastly outperform a typical sparse, tri-diagonal method, the finite difference method. Further, in the fluxonium, both phase and charge number DVRs of various grid sizes achieve more accurate and efficient level convergence than the harmonic oscillator basis. In the more restrictive case of the transmon, where periodicity limits the usage of DVRs to the charge basis and an analogous discrete phase basis, we still find that the phase DVR can match the performance of this typical representation. We additionally show that both the infinite (the traditional sinc DVR) and finite (the truncated sinc DVR) descriptions perform similarly, a promising feature that may prove valuable in the context of time evolution via trotterization, where the conjugate variables of the truncated sinc DVR are related by a discrete Fourier transform.

While our work offers a potentially more efficient and accurate numerical representation of superconducting circuits, important questions remain regarding the applicability of these representations (and others) to larger-scale systems. Sparsity is likely an important feature for multi-qubit simulations, where smaller and/or truncated state spaces are invaluable for handling numerous degrees of freedom. Other sparse approaches, such as wavelets or higher-order finite difference methods, may be able to leverage their sparsity to represent these larger-scale systems while still reaching decoherence accuracy. Moving forward, further investigation into approaches that could combine the level of accuracy provided by DVRs with the sparsity of alternative methods may offer promising solutions for efficiently modeling larger-scale quantum systems and expanding the ever-growing capabilities of software tools for superconducting circuits.

\begin{acknowledgements}

We thank Eite Tiesinga, Jose Aumentado, and Michael Gullans for helpful discussions and feedback.
B.R. is supported by the U.S. DOE Office of Science, Office of High Energy Physics, QuantISED program (under FWP ERKAP63).

\end{acknowledgements}

\appendix

\section{Discrete variable representations}
\label{DVRbkgrd}

In this appendix, we provide an introduction to discrete variable representations (DVRs), including derivations of their required properties and clarifications on their limitations in numerical analysis. We outline the construction and show the associated basis states of a classic DVR, the orthogonal polynomial DVR, as well as present the relevant details for building the sinc DVRs utilized in the main text.

A DVR~\cite{DVRvargensinc,genDVR1,genDVR2} is an orthonormal and complete basis set that is generated by taking a discrete set of points from a system's degrees of freedom and projecting them onto another orthonormal and complete basis set. This yields a basis in which individual basis functions are localized around their associated grid points. Different DVRs are constructed by considering different orthonormal and complete function sets to project onto. Explicitly, we construct a DVR by combining a projector $\hat{P}$ and a set of grid points $\lbrace x_\alpha \rbrace$ such that the resulting states,
\begin{equation}
\label{eq:DVRstate}
\ket{\psi_{\alpha,x}} = \frac{1}{\sqrt{c_\alpha}} \hat{P} \ket{x_\alpha} \enspace ,
\end{equation}
are orthonormal and complete. That is to say,
\begin{equation}
\label{eq:DVRconditions}
\begin{split}
&\bra{\psi_{\alpha,x}}\ket{\psi_{\beta,x}} = \delta_{\alpha \beta} \\
&\sum_{\alpha} \ket{\psi_{\alpha,x}} \bra{\psi_{\alpha,x}} = \hat{P} \enspace .
\end{split}
\end{equation}
Here, $c_\alpha$ is a normalization constant while the indices $\alpha$ and $x$ serve to label each DVR state with its grid point $\alpha$ and the discretized variable $x$. We note that within the subspace defined by $\hat{P}$, the completeness relation in Eq.~(\ref{eq:DVRconditions}) is the identity, as expected.

The projector $\hat{P}$ is assembled from some orthonormal and complete function set $\lbrace f_k(x) \rbrace$, where $k$ may index each function either discretely or continuously. For a function set with discretely-indexed basis functions and orthonormality and completeness relations 
\begin{equation}
\label{eq:basisconditions}
\begin{split}
&\int_a^b dx \ f_k^*(x) f_j(x)  = \delta_{kj}\\
&\sum_{k} f_k^*(x) f_k(x^\prime) = \delta(x-x^\prime) \enspace ,
\end{split}
\end{equation}
the projector takes the form
\begin{equation}
\label{eq:DP}
\hat{P} = \sum_{k=1}^M \ket{f_k} \bra{f_k} \enspace ,
\end{equation}
where $N$ indicates the subspace defined by the projector and as usual, $f_k(x) = \bra{x}\ket{f_k}$. If instead the function set is continuously indexed [i.e., $\delta_{kj} \rightarrow \delta(k-j)$ and $\sum_k \rightarrow \int dk$ in Eq.~(\ref{eq:basisconditions})], the projection operator takes the form 
\begin{equation}
\label{eq:CP}
\hat{P} = \int_{-k_{\rm max}}^{k_{\rm max}} dk \ket{f_k} \bra{f_k}  \enspace ,
\end{equation}
where now, the subspace of the continuous index $k$ is defined by the boundary values $\pm k_{\rm max}$.

In both instances, this formulation of a DVR and the conditions in Eq.~(\ref{eq:DVRconditions}) highlight important features and requirements of this representation. We identify an essential aspect of a DVR's basis functions by re-expressing the orthonormality condition in Eq.~(\ref{eq:DVRconditions}),
\begin{equation}
\label{eq:interpolation}
\begin{split}
\bra{\psi_{\alpha,x}}\ket{\psi_{\beta,x}} &= \frac{1}{\sqrt{c_\alpha^* c_\beta}} \bra{x_\alpha} \hat{P}^\dagger \hat{P} \ket{x_\beta} \\
&= \frac{1}{\sqrt{c_\alpha^* c_\beta}} \bra{x_\alpha} \hat{P} \ket{x_\beta}\\
&= \frac{1}{\sqrt{c_\alpha^*}} \psi_{\beta,x}(x_\alpha) = \delta_{\alpha \beta} \enspace ,
\end{split}
\end{equation}
where we exploit the projection operator property $\hat{P} = \hat{P}^\dagger = \hat{P}^2$ and define $\bra{x_\alpha}\ket{\psi_{\beta,x}} = \psi_{\beta,x}(x_\alpha)$. Namely, each basis function, by definition, must vanish at every grid point other than its own. Further, by combining Eq.~(\ref{eq:DVRstate}) and either Eq.~(\ref{eq:DP}) or (\ref{eq:CP}) with the conditions in Eq.~(\ref{eq:DVRconditions}), we see that the functions $\lbrace f_k(x) \rbrace$ used to construct $\hat{P}$ must satisfy both an exact quadrature rule that discretizes their orthonormality, 
\begin{equation}
\label{eq:quadrature}
\sum_\alpha \frac{1}{|c_\alpha|} f_k^*(x_\alpha) f_{k^\prime}(x_\alpha) = \begin{cases}
\delta_{kk^\prime} \\
\delta(k-k^\prime)
\end{cases}  \enspace ,
\end{equation}
where we can identify the constant $|c_\alpha|$ with the weights of our desired quadrature rule, as well as a discretized completeness relation within the projector's subspace at the grid points $\lbrace x_\alpha \rbrace$:
\begin{equation}
\label{eq:Dcomplete}
\sqrt{c_\alpha^* c_\beta} \delta_{\alpha \beta} = \begin{cases}
\sum_{k=1}^N f_k(x_\alpha) f_k^*(x_\beta) \\
\\
\int_{-k_{\rm max}}^{k_{\rm max}} dk f_k(x_\alpha) f_k^*(x_\beta) 
\end{cases} \enspace .
\end{equation}
Thus in practice, constructing a DVR consists of choosing a function set $\lbrace f_k(x) \rbrace$, projector $\hat{P}$, constant $c_\alpha$, and grid $\lbrace x_\alpha \rbrace$ such that Eqs.~(\ref{eq:quadrature}) and (\ref{eq:Dcomplete}) are satisfied.

The final crucial (and most practical) characteristic of a DVR is the diagonal approximation, where we assume the matrix form of operators that are a function of the discretized variable are diagonal in the DVR's basis, evaluated at each point on the grid:
\begin{equation}
\label{eq:DA}
\bra{\psi_{\alpha,x}} V(\hat{x}) \ket{\psi_{\beta,x}} \approx V(x_\alpha) \delta_{\alpha \beta} \enspace .
\end{equation}
This property hinges on the quadrature rule in Eq.~(\ref{eq:quadrature}). One can readily confirm that Eq.~(\ref{eq:DA}) only holds if we approximate the integral,
\begin{equation}
\begin{split}
\bra{f_k} V(\hat{x})\ket{f_{k^\prime}} &= \int_a^b dx \ f_k^*(x) V(x) f_{k^\prime}(x) \\
&\approx \sum_\alpha \frac{1}{|c_\alpha|} f_k^*(x_\alpha) V(x_\alpha) f_{k^\prime}(x_\alpha)
\end{split} \enspace ,
\end{equation}
using a quadrature rule based on Eq.~(\ref{eq:quadrature}). This approximation introduces a potential quadrature error that we interpret as a consequence of the fact that $V(\hat{x}) \ket{\psi_{\beta,x}}$ may not necessarily lie inside the subspace defined by $\hat{P}$. Importantly, these errors are not guaranteed to be positive. Thus, when DVRs are used to model Hamiltonians, this source of error may not necessarily increase the energy eigenvalues. Since the estimated eigenvalues may be below their actual values, DVRs are non-variational basis representations.

We now consider the details for the construction of three different DVRs. For illustrative purposes, we first discuss a classic DVR based on real orthogonal polynomials. Then, we present the derivation and details of the two DVRs utilized in the main text. These are constructed using variations of the Fourier basis, with the traditional `sinc DVR' using the continuous Fourier basis and the truncated sinc DVR using the partially discrete Fourier basis.

\subsection{Orthogonal polynomial DVRs}

Real orthogonal polynomials~\cite{orthoAS} are characterized by their orthogonality relation,
\begin{equation}
\int_a^b dx \ w(x) p_m(x) p_n(x) = \tilde{c}_n \delta_{mn} \enspace ,
\end{equation}
where $p_m(x)$ is the $m^{\rm th}$-degree polynomial on the interval~$[a,b]$ with coefficient $\tilde{c}_n$ and weight function $w(x)$. Distinct families of orthogonal polynomials (e.g., Hermite polynomials, Legendre polynomials, etc.) are characterized by their different intervals, coefficients, and weight functions. Importantly, via Gaussian quadrature~\cite{gauss1,gauss2}, this orthogonality relation can be rewritten using an exact quadrature rule,
\begin{equation}
\int_a^b dx \ w(x) p_m(x) p_n(x) = \sum_{i=1}^M \tilde{w}_i p_m(x_i) p_n(x_i) \enspace ,
\end{equation}
where $\lbrace x_i \rbrace$ are the zeros of $p_M(x)$ and $\tilde{w}_i$ are the weights,
\begin{equation}
\tilde{w}_i = \int_a^b dx \ L_i(x) w(x) \enspace ,
\end{equation}
with
\begin{equation}
L_i(x) = \prod_{\substack{j=1 \\ j \neq i}}^M \frac{x-x_j}{x_i-x_j} \enspace .
\end{equation}
This quadrature rule is exact for polynomial functions with a degree less than or equal to $2M-1$, i.e., the degree of the function $p_m(x) p_n(x)$ such that $m+n \leq 2M-1$.

This quadrature rule yields a straightforward path to creating a DVR based on orthogonal polynomials. With a projector assembled from the functions 
\begin{equation}
f_k(x) = \sqrt{\frac{w(x)}{\tilde{c}_k}} p_k(x) \enspace ,
\end{equation}
we project onto a subspace of up to $M-1$ polynomials, thus ensuring the exact quadrature rule is satisfied. Then, with a grid of points corresponding to the zeros $p_M(x)$ and the coefficients
\begin{equation}
c_\alpha = \frac{w(x_\alpha)}{\tilde{w}_\alpha} \enspace ,
\end{equation}
the conditions outlined in Eqs.~(\ref{eq:DVRconditions}), (\ref{eq:quadrature}), and (\ref{eq:Dcomplete}) are satisfied and we find the DVR's basis states to be
\begin{equation}
\ket{\psi_{\alpha,x}} = \sum_{k=1}^{M-1} \sqrt{\frac{\tilde{w}_\alpha}{\tilde{c}_k}} p_k(x_\alpha) \ket{f_k} \enspace .
\end{equation}
We see that a set of orthogonal polynomials may always be used to create a DVR, since they are orthonormal and complete on points that correspond to the zeros of the first polynomial excluded from the subspace.

\subsection{The traditional sinc DVR}
\label{tradsincDER}

The usual sinc DVR~\cite{chemphyssinc,chemphysgensinc,DVRvargensinc,periodicDVR} is formed by using the continuous Fourier basis to project onto a discrete set of points, exploiting the orthonormality and completeness of both the continuous Fourier basis as well as its partially discrete representation. The continuous Fourier basis~\cite{arfken},
\begin{equation}
\label{eq:CFB}
f(x,p) = \frac{e^{i p x / \hbar}}{\sqrt{2 \pi \hbar}} \enspace , 
\end{equation}
is used in the Fourier transform to continuously project a function in real space onto its conjugate space or vice versa, which provides the continuous orthonormality and completeness relations,
\begin{equation}
\label{eq:orthCFB}
\begin{split}
\delta(p-p^\prime) = \frac{1}{2 \pi \hbar} \int_{-\infty}^\infty dx \ e^{-i (p-p
^\prime) x/ \hbar} \\
\delta(x-x^\prime) = \frac{1}{2 \pi \hbar} \int_{-\infty}^\infty dp \ e^{i p (x-x^\prime) / \hbar} \enspace ,
\end{split}
\end{equation}
where both the real-space variable $x$ and the conjugate variable $p$ are continuous with infinite extent.

We can leverage the continuous Fourier basis to construct a DVR by noting the orthonormality and completeness of the partially discrete Fourier basis, as utilized in the complex Fourier series expansion of a function in real space or conjugate space. A Fourier series expansion uses this partially discrete representation to project a continuous function onto an infinite number of discrete points, representing this function discretely in one space and continuously, but on a finite interval, in the other. The Fourier series expansion of a real-space function provides the orthonormality and completeness of the partially discrete Fourier basis in conjugate space,
\begin{equation}
\label{eq:FSrelationP}
\begin{split}
\delta(p-p^\prime) &= \frac{1}{2 p_{\rm max}} \sum_{n = -\infty}^{\infty} e^{-i 2 \pi n (p-p^\prime)/2 p_{\rm max}} \\
\delta_{mn} &= \frac{1}{2 p_{\rm max}} \int_{-p_{\rm max}}^{p_{\rm max}} dp \ e^{-i 2 \pi (m-n) p / 2 p_{\rm max}} \enspace ,
\end{split} 
\end{equation}
which we identify as precisely the exact quadrature and completeness relations [Eqs.~(\ref{eq:quadrature}) and (\ref{eq:Dcomplete})] needed to construct a DVR. Similarly, the Fourier series expansion of a conjugate-space function provides the orthonormality and completeness of the partially discrete Fourier basis in real space,
\begin{equation}
\label{eq:FSrelationX}
\begin{split}
\delta(x-x^\prime) &= \frac{1}{2 x_{\rm max}} \sum_{n = -\infty}^{\infty} e^{i 2 \pi n (x-x^\prime)/2 x_{\rm max}} \\
\delta_{mn} &= \frac{1}{2 x_{\rm max}} \int_{-x_{\rm max}}^{x_{\rm max}} dx \ e^{i 2 \pi (m-n) x / 2 x_{\rm max}} \enspace .
\end{split} 
\end{equation}
Thus, a projector assembled from the continuous Fourier basis [Eq.~(\ref{eq:CFB})] can be used in two ways to create a DVR, either discretizing the real-space variable $x$ with the conjugate variable $p$ continuous and bounded or discretizing $p$ with $x$ continuous and bounded.

For a DVR that discretizes $x$, we use a projector assembled from the real-space functions $\bra{x}\ket{f_p} = f(x,p)$, which project onto a continuous-$p$ subspace bounded by $\pm p_{\rm max}$ [as in Eq.~(\ref{eq:CP})]. Examining Eqs.~(\ref{eq:orthCFB}) and (\ref{eq:FSrelationP}), we see that this projector can be used to build a DVR when combined with an infinite, uniform grid of points $x_\alpha = \alpha \Delta x$, spaced according to $\Delta x = \pi \hbar / p_{\rm max}$, and a coefficient $c_\alpha = 1/ \Delta x = p_{\rm max} / \pi \hbar$. The resulting DVR states take a form analogous to Eq.~(\ref{eq:tradsincTH}) in the main text, which instead considers phase $\theta$ and charge number $N$ as the two (dimensionless) conjugate degrees of freedom.

Analogously, for a DVR that discretizes $p$, we instead use a projector assembled from the conjugate-space functions $\bra{p}\ket{f_x} = f(x,p)^*$, which project onto a continuous-$x$ subspace bounded by $\pm x_{\rm max}$. We note that the sign change in the projector's functions serves to maintain the usual relationship between the conjugate spaces. Examining Eqs.~(\ref{eq:orthCFB}) and (\ref{eq:FSrelationX}), we see that we can once again construct a DVR, now with the infinite, uniform grid $p_\alpha = \alpha \Delta p$, spaced according to $\Delta p = \pi \hbar/ x_{\rm max}$, and a coefficient $c_\alpha = 1/ \Delta p = x_{\rm max} / \pi \hbar$. The resulting DVR in this case is given in Eq.~(\ref{eq:tradsincN}) in the main text, again considering phase $\theta$ and charge number $N$ as the two conjugate degrees of freedom.

We note that due to the required infinite number of basis states (and infinite extent of the discretized variable), in practice, these DVRs are subject to a truncation error associated with the number of DVR states included in the state space. Typically, we proceed with the infinite state space description of the states and operators, recognizing that this error is minimized with a sufficiently large state space.

\subsection{The truncated sinc DVR}
\label{truncsincDER}

The notion of truncating the infinite traditional sinc DVR naturally leads to the introduction of the second DVR one can construct using the Fourier basis. In this case, the DVR's basis is instead finite-sized by definition, requiring no additional truncation for numerical implementation. As previously indicated by Eqs.~(\ref{eq:FSrelationP}) and (\ref{eq:FSrelationX}), the partially discrete representation utilized in the Fourier series is also orthonormal and complete, but on a finite interval. Written explicitly, the partially discrete real-space functions 
\begin{equation}
\label{eq:xcon}
f_n(x) = \frac{e^{i p_n x / \hbar}}{\sqrt{2 x_{\rm max}}} \enspace ,
\end{equation}
are orthonormal and complete on the interval $x \in [-x_{\rm max}, x_{\rm max}]$, with $p_n = n \Delta p$, $x_{\rm max} = \pi \hbar /\Delta p$, and $n$ integer-valued on the range $[-\infty,\infty]$. Analogously, the partially discrete conjugate-space functions
\begin{equation}
\label{eq:pcon}
g_n(p) = \frac{e^{-i p x_n / \hbar}}{\sqrt{2 p_{\rm max}}} \enspace ,
\end{equation}
are orthonormal and complete on the interval $p \in [-p_{\rm max}, p_{\rm max}]$, with $x_n = n \Delta x$, $p_{\rm max} = \pi \hbar/\Delta x$, and $n$ integer-valued on the range $[-\infty,\infty]$. We now consider the use of these functions for the projection operator $\hat{P}$ in order to construct a DVR.

The exact quadrature and discrete completeness relations required to build a DVR in this case are enabled by the discrete Fourier transform~\cite{arfken}, which defines the orthonormality and completeness of the fully discrete representation of the Fourier basis, given by
\begin{equation}
\label{eq:DOC}
\delta_{jk} = \frac{1}{2 M +1} \sum_{n = -M}^{M} e^{i 2 \pi n (j-k)/2M+1} \enspace ,
\end{equation}
where $|j-k| < 2M+1$. In this instance, the creation of the DVR amounts to discretizing the remaining continuous and bounded variable in Eqs.~(\ref{eq:xcon}) or (\ref{eq:pcon}), yielding a basis set which is finite-sized and discrete in both the real-space and conjugate-space variables. As we show in the main text, while these two DVRs are similar in this manner, the resulting basis states are distinct, specifically with regard to periodicity.

Using a projection operator assembled from the functions in Eq.~(\ref{eq:xcon}), we proceed with a discretization of the real-space variable $x$ by comparing Eqs.~(\ref{eq:FSrelationX}) and (\ref{eq:DOC}). Upon examination, we can construct a DVR provided we project into a finite subspace of $2M+1$ states [as in Eq.~(\ref{eq:DP})], where $n \in [-M,M]$, using the constant $c_\alpha = 2M+1/2 x_{\rm max}$ and a grid of points $x_\alpha = \alpha \Delta x$, which is finite-sized ($\alpha \in [-M,M]$) and spaced according to $\Delta x \Delta p = 2 \pi \hbar / 2M+1$. Note that from the original basis functions of the projector, we have that $p_n = n \Delta p$ and $x_{\rm max} = \pi \hbar/\Delta p$, allowing us to re-express the constant as $c_\alpha = (2 M+1) \Delta p / 2\pi \hbar = 1/\Delta x$. 
In the main text, the resulting DVR states are given in terms of $\theta$ and $N$ in Eq.~(\ref{eq:truncsincTH}).

This procedure is carried out analogously for a projection operator assembled from the functions in Eq.~(\ref{eq:pcon}). Here, we discretize the conjugate variable $p$ to construct a DVR, comparing Eqs.~(\ref{eq:FSrelationP}) and (\ref{eq:DOC}). In this case, we must project into a finite subspace of $2M+1$ states, where $n \in [-M,M]$, with the constant $c_\alpha = 2M+1/2 p_{\rm max}$ and a grid of points $p_\alpha = \alpha \Delta p$, which is finite-sized ($\alpha \in [-M,M]$) and spaced according to $\Delta x \Delta p = 2 \pi \hbar/ 2M+1$. Note again that from the original basis functions of the projector, we have that $x_n = n \Delta x$ and $p_{\rm max} = \pi \hbar/\Delta x$, allowing us to re-express the constant as $c_\alpha = (2 M+1) \Delta x / 2\pi \hbar = 1/\Delta p$. In this case, we find DVR states with a form equivalent to that given in Eq.~(\ref{eq:truncsincN}).

\section{Details for the LC oscillator example}
\label{LCextra}

\begin{figure*}
\centering
\includegraphics[width= 0.96\textwidth]{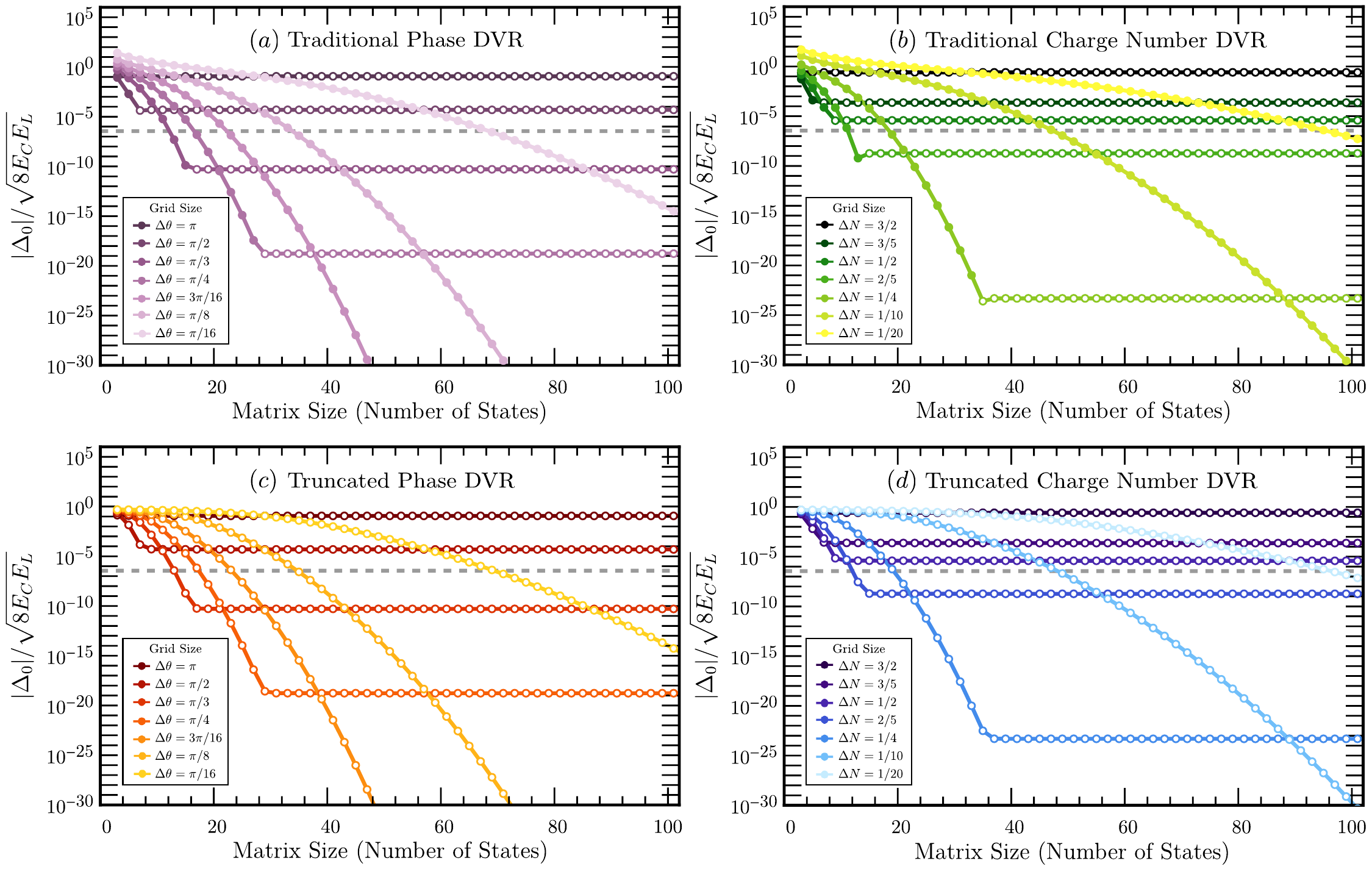}
\caption{The absolute value of the energy difference $|\Delta_0|$ as function of matrix size in the ground state of the LC oscillator, in all four DVRs presented in Section~\ref{S2}: (a) the traditional phase DVR, (b) the traditional charge number DVR, (c) the truncated phase DVR, and (d) the truncated charge number DVR. All are scaled according to the LC frequency. Decoherence accuracy (also scaled according to the LC frequency) is indicated with a horizontal gray dashed line. Open data points indicate where $\Delta_0 <0$, in order to note and highlight the non-variational nature of DVRs.}
\label{fig:LCcurves}
\end{figure*}

In this appendix, we provide additional details for the analysis performed for the LC oscillator presented in Section~\ref{LCoscillator}. To generate the results in Fig.~\ref{fig:LCmetrics}, a numerical analysis for each of the four DVRs is carried out for a collection of grid sizes and matrix sizes, generating the energy eigenvalues associated with each pairing. For the traditional and truncated charge number DVRs, this includes nineteen grid sizes between $\Delta N = 1/20$ and $\Delta N = 2$, the majority of which are visible in Fig.~\ref{fig:LCmetrics}. Similarly, for the traditional and truncated phase DVRs, this includes eighteen grid sizes between $\Delta \theta = \pi / 64$ and $\Delta \theta = 3 \pi$. With the differing scale of these ranges, which we attribute to the physical nature of each variable, not all points are visible in Fig.~\ref{fig:LCmetrics}. In all cases, up to 150 digits of precision are maintained, far beyond any precision of interest, so as to not affect visible results. We set $E_L/h = E_C/h = 1$ GHz for simplicity, however, it is ultimately scaled away by considering $|\Delta_0|/\sqrt{8 E_C E_L}$. Additionally, matrix sizes out to 301 basis states are considered to capture the surpassing of decoherence accuracy and saturation behavior. In Fig.~\ref{fig:LCcurves}, we include a sampling of these grid sizes in all four DVRs, restricting the range of accuracy and matrix sizes to those most relevant.

Comparing the traditional and truncated DVRs (Figs.~\ref{fig:LCcurves}a and \ref{fig:LCcurves}c or Figs.~\ref{fig:LCcurves}b and \ref{fig:LCcurves}d), we note the improved accuracy of the truncated DVRs at very small matrix sizes. As expected from the theory, this regime is where the required numerical truncation of the traditional DVR is most impactful, as we are approximating matrix elements using infinite basis-based expressions in very small matrices. This discrepancy, while dependent on grid size, becomes negligible upon reaching matrix sizes corresponding to saturation precision. We also note the non-variational behavior evident in all four DVRs shown in Fig.~\ref{fig:LCcurves}, as evidenced by the regions where $\Delta_0 <0$. While we can confirm that the energies themselves ($E_0^{\rm DVR}$) monotonically decrease with increasing matrix size, as anticipated, their difference from the true energy, $\Delta_0$, is not exclusively positive. In the traditional DVRs, we find that while the DVRs' energies do approach from above the true energy, they ultimately converge below it. Upon passing below the true energy, it occasionally leads to a sharp feature, like that in Fig.~\ref{fig:LCcurves}b for $\Delta N = 2/5$, where the saturation precision below the true energy is actually larger than that given by the final matrix size for which $\Delta_0 >0$. In contrast, in the truncated case, the energies exclusively converge to the true energy from below.

\begin{figure*}
\centering
\includegraphics[width= 0.96\textwidth]{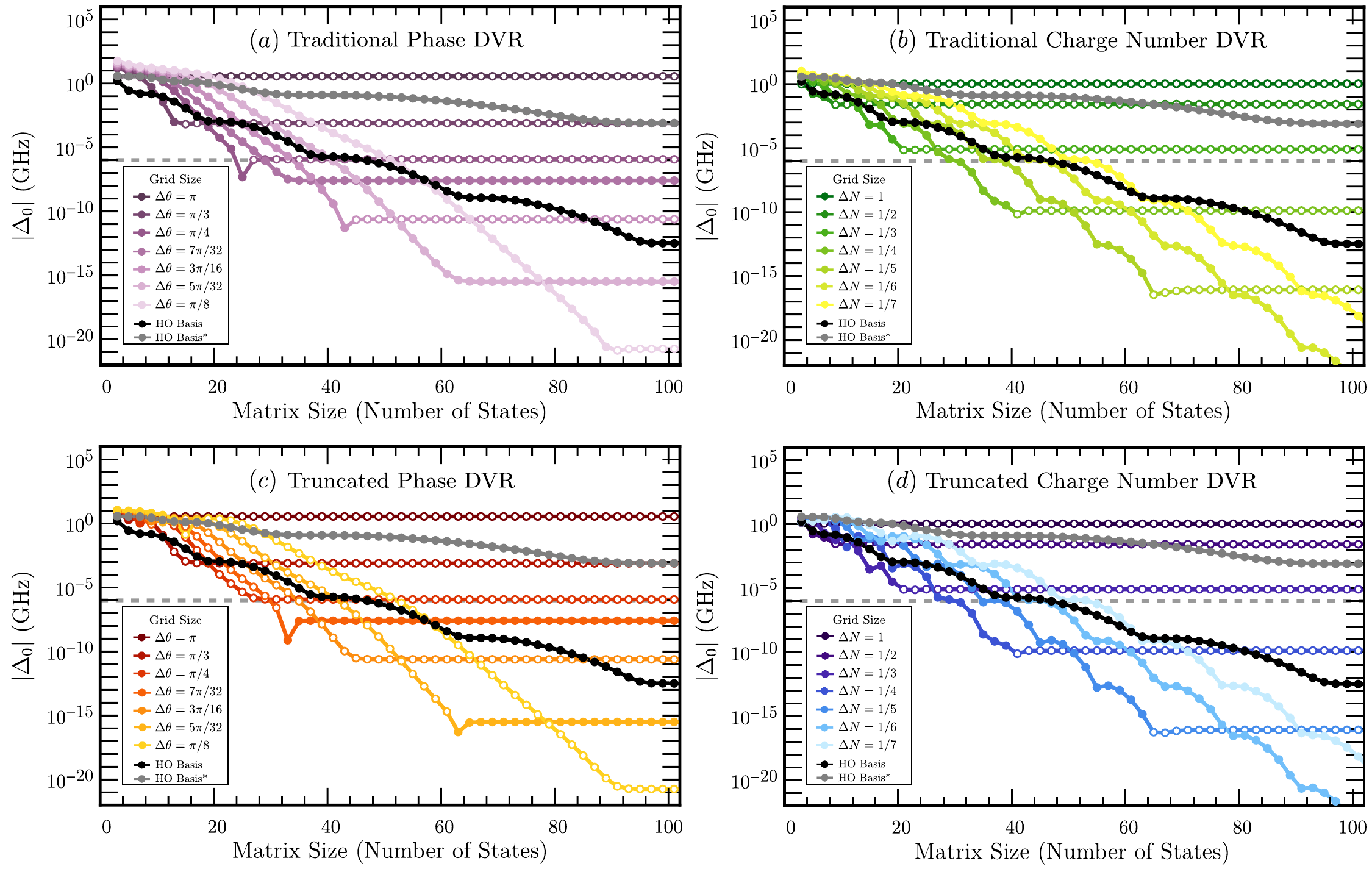}
\caption{The absolute value of the energy difference $|\Delta_0|$ as function of matrix size in the ground state of the fluxonium, in all four DVRs presented in Section~\ref{S2}: (a) the traditional phase DVR, (b) the traditional charge number DVR, (c) the truncated phase DVR, and (d) the truncated charge number DVR. Decoherence accuracy is indicated with a horizontal gray dashed line. Open data points indicate where $\Delta_0 <0$, in order to note and highlight the non-variational nature of DVRs. Included in each subplot are two curves indicating the performance of the harmonic oscillator basis using two different length scales. The black curve represents the standard approach using a length scale corresponding to the LC frequency $\omega_{\rm LC} = \sqrt{8 E_C E_L} / h$. The gray curve modifies the length scale to instead use the plasma frequency $\omega_{\rm PL} = \sqrt{8 E_C E_J} / h$. Parameters for this plot are $E_C/h = 2.5$ GHz, $E_L/h = 0.5$ GHz, $E_J/h = 10$ GHz, and $\mathcal{A} = 1/2$.}
\label{fig:FLcurves}
\end{figure*}

\section{Details for the fluxonium example}
\label{FLextra}

In this appendix, we provide additional details for the analysis performed for the fluxonium presented in Section~\ref{Fluxonium}. To generate the results in Fig.~\ref{fig:FLmetrics}, a numerical analysis for each of the four DVRs is carried out for a collection of grid sizes and matrix sizes, generating the energy eigenvalues associated with each pairing. For the traditional and truncated charge number DVRs, we consider the subset of grid sizes where $1/\Delta N$ is an integer, in order to consider a simplified matrix form for the cosine operator, as we discuss in Section~\ref{OpRep}. Thus, we use the grid sizes $\Delta N = 1/n$ with $n= 1,...,15$, generating fifteen grid sizes, the majority of which are visible in Fig.~\ref{fig:FLmetrics}. Identical to the LC oscillator, for the traditional and truncated phase DVRs, we consider eighteen grid sizes between $\Delta \theta = \pi / 64$ and $\Delta \theta = 3 \pi$. With the differing scale of these ranges, which we attribute to the physical nature of each variable, not all points are visible in Fig.~\ref{fig:FLmetrics}. In all cases, up to 150 digits of precision are maintained, far beyond any precision of interest, so as to not affect visible results. We set the energy scales and flux penetration of the circuit to reasonable values~\cite{fluxonium,fluxoniumEX,fluxcoherence,fluxcoherence2} for the fluxonium --- $E_C/h = 2.5$ GHz, $E_L/h = 0.5$ GHz, $E_J/h = 10$ GHz, and $\mathcal{A} = 1/2$. Additionally, matrix sizes out to 301 basis states are considered to capture the surpassing of decoherence accuracy and saturation behavior. In Fig.~\ref{fig:FLcurves}, we include a sampling of these grid sizes in all four DVRs, restricting the range of accuracy and matrix sizes to those most relevant.

Included in Fig.~\ref{fig:FLcurves} are two curves indicating the performance of the harmonic oscillator basis using two different length scales. We define the harmonic oscillator basis in terms of the raising and lowering operators $\hat{a}^\dagger$ and $\hat{a}$ as well as an associated length scale $\theta_0$, where
\begin{equation}
\label{eq:HObasis1}
\hat{\theta} = \frac{\theta_0}{\sqrt{2}} ( \hat{a}^\dagger + \hat{a} )
\end{equation}
and
\begin{equation}
\label{eq:HObasis2}
\hat{N} = \frac{i}{\sqrt{2} \theta_0} ( \hat{a}^\dagger - \hat{a} ) \enspace .
\end{equation}
Using dimensional analysis in the LC oscillator's Hamiltonian [Eq.~(\ref{eq:LCham})], this length scale is expressed generally as $\theta_0 = (\hbar \omega/E_L)^{1/2}$, with $\omega$ the frequency associated with the desired length scale. In the standard harmonic oscillator basis, $\omega$ is taken to be the LC frequency $\omega_{\rm LC} = \sqrt{8 E_C E_L} / h$, yielding the usual length scale $\theta_0 = (8 E_C/E_L)^{1/4}$. This approach considers a length scale appropriate for the first two terms of the fluxonium Hamiltonian [Eq.~(\ref{eq:FLham})], diagonalizing these terms, but still requiring a matrix exponential to be evaluated in the final third term. Alternatively, one can consider using the other relevant frequency scale in the fluxonium, the plasma frequency $\omega_{\rm PL} = \sqrt{8 E_C E_J} / h$, which is appropriate for the first and last terms of the fluxonium Hamiltonian [Eq.~(\ref{eq:FLham})]. This results in the length scale $\theta_0 = (\sqrt{8 E_C E_J}/E_L)^{1/2}$. In both cases, due to the required numerical evaluation of a matrix exponential, each matrix size considered in Fig.~\ref{fig:FLcurves} is a sub-matrix of an original matrix consisting of 1001 basis states.

Upon comparison, we find that the standard harmonic oscillator basis greatly outperforms that with a modified length scale, as indicated by its more rapid descent to decoherence accuracy. The standard harmonic oscillator basis surpasses decoherence accuracy at 47 basis states, as shown in Fig.~\ref{fig:FLmetrics}a, while the harmonic oscillator basis with a length scale modified according to $\omega_{\rm PL}$ requires 209 basis states, outside the visible range in both Fig.~\ref{fig:FLmetrics} and Fig.~\ref{fig:FLcurves}. However, more promising is the performance of all four DVRs in comparison with the standard harmonic oscillator basis. There are various grid sizes for which level convergence is achieved at higher accuracy and smaller matrix sizes when compared to the standard method. While Fig.~\ref{fig:FLmetrics} specifically highlights decoherence accuracy and saturation precision, including the improvements in reaching decoherence accuracy that all four DVRs provide, Fig.~\ref{fig:FLcurves} displays a complete picture of performance, specifically with regard to matrix size and its associated precision. Namely, that the DVRs outlined here, for small enough grid sizes, appear capable of outperforming the standard numerical approach with regard to accuracy and matrix size, even beyond the decoherence-accurate level.

Comparison between the traditional and truncated DVRs (Figs.~\ref{fig:FLcurves}a and \ref{fig:FLcurves}c or Figs.~\ref{fig:FLcurves}b and \ref{fig:FLcurves}d) yields a similar conclusion as in the LC oscillator. We find an improvement in the accuracy of the truncated DVRs over the traditional DVRs at small matrix sizes, which diminishes as the matrix size increases, and becomes negligible as the matrix sizes corresponding to saturation precision are reached. The non-variational behavior of each DVR is similarly on display in this example, albeit with more variety. For the charge number DVRs, we generally observe similar behavior as in the LC oscillator, with energies that converge from above to a value below the reference energy. However, for the phase DVRs, we find grid sizes that converge to values above or below the reference energy, and not exclusively from above. As in the LC oscillator, this transitional behavior from $\Delta_0 > 0$ to $\Delta_0 <0$ or vice versa is occasionally marked by sharp features in the plots in Fig.~\ref{fig:FLcurves}.

\section{The finite difference method}
\label{FDM}

In this appendix, we outline and derive one of the most well-known discretization methods, the finite difference (FD) method~\cite{compphys,FDrev,Fornberg1988,Vich1981}, which we use for comparison with DVRs in the main text.

\subsection{Lowest order FD method}

In a continuous position basis, the Schr$\ddot{\text{o}}$dinger equation takes the usual form
\begin{equation}
-\frac{\hbar^2}{2 m} \frac{\partial^2 \psi_n(x)}{\partial x^2} + V(x) \psi_n(x) = E_n \psi_n(x) \enspace .
\label{eq:usualS}
\end{equation}
By discretizing this system over a uniform grid of points $x_i$ separated by $\Delta x$, such that $x_i = i \Delta x$ for some total number of points $i \in [-N,N]$, we can evaluate and solve Eq.~(\ref{eq:usualS}) at each point. This requires we consider the second derivative evaluated at each grid point, $\frac{\partial^2 \psi_n(x_i)}{\partial x^2}$, which we can approximate using a centered FD approximation, 
\begin{equation}
\frac{\partial^2 \psi_n(x_i)}{\partial x^2} \approx \frac{\psi_n(x_{i-1})-2\psi_n(x_i)+\psi_n(x_{i+1})}{\Delta x^2} \enspace .
\end{equation}
The discretized Schrodinger equation then takes the form
\begin{equation}
\begin{split}
-\frac{\hbar^2}{2 m} &\frac{\psi_n(x_{i-1})-2\psi_n(x_i) +\psi_n(x_{i+1})}{\Delta x^2} \\
&+ V(x_i) \psi_n(x_i) = E_n \psi_n(x_i) \enspace .
\end{split}
\end{equation}
This equation represents a system of equations for the $2N+1$ points in the grid, which we can write in matrix form as 
\begin{widetext}
\begin{equation}
\begin{split}
\frac{-\hbar^2}{2m} \frac{1}{\Delta x^2} \begin{bmatrix}
\textrm{-}2&1& \ & \  \\
1&\textrm{-}2&1& \ & \ \\
 \ &1&\textrm{-}2 \\
 \ & \ & &  \ddots \\
 \ & \ & \ & \ &\textrm{-}2& 1& \ \\
 \ & \ & \ & \ &1&\textrm{-}2&1 \\
  \ & \ & \ & \ & \ &1&\textrm{-}2
\end{bmatrix} & \begin{bmatrix}
\psi_n(x_{\textrm{-}N}) \\
\vdots \\
\psi_n(x_0) \\
\vdots \\
\psi_n(x_{N})
\end{bmatrix} + \frac{-\hbar^2}{2m} \frac{1}{\Delta x^2} \begin{bmatrix}
\psi_n(x_{\textrm{-}N\textrm{-}1}) \\
0 \\
\vdots \\
0 \\
\psi_n(x_{N+1})
\end{bmatrix} \\
&+ \begin{bmatrix}
V(x_{\textrm{-}N})& \ & \ & \  \\
 \ & \hspace{-1em} \ddots& \ & \ & \ \\
 \ & \ & & \hspace{-1em} V(x_0) \\
 \ & \ & \ & \ & \ & \hspace{-1em} \ddots \\
  \ & \ & \ & \ & \ & \ & \hspace{-1em} V(x_{N})
\end{bmatrix} \begin{bmatrix}
\psi_n(x_{\textrm{-}N}) \\
\vdots \\
\psi_n(x_0) \\
\vdots \\
\psi_n(x_{N})
\end{bmatrix} = E_n \begin{bmatrix}
\psi_n(x_{\textrm{-}N}) \\
\vdots \\
\psi_n(x_0) \\
\vdots \\
\psi_n(x_{N})
\end{bmatrix} \enspace .
\end{split}
\label{eq:MATSeq}
\end{equation}
\end{widetext}
The second term in Eq.~(\ref{eq:MATSeq}) is understood as the boundary condition term, as it references points outside of the pre-defined grid. By explicitly including this term, we can consider the cases of either a bounded or periodic potential $V(x)$, as in the examples in Section~\ref{S3}.

If the potential $V(x)$ is bounded such that $\psi_n(x) \rightarrow 0$ as $x \rightarrow \pm \infty$, we take the boundary terms corresponding to $\psi_n(x_{-N-1})$ and $\psi_n(x_{N+1})$ to be zero. This corresponds to letting $V(x) \rightarrow \infty$ at the edges of our pre-defined grid, which should be a reasonable approximation so long as $N \Delta x$ is large enough to fully capture an eigenstate of interest. This leaves a straightforward eigenvalue problem to solve,
\begin{widetext}
\begin{equation}
\label{eq:FDbounded}
\left( \frac{-\hbar^2}{2m} \frac{1}{\Delta x^2} \begin{bmatrix}
\textrm{-}2&1& \ & \  \\
1&\textrm{-}2&1& \ & \ \\
 \ &1&\textrm{-}2 \\
 \ & \ & \ & \ddots \\ 
 \ & \ & \ & \ &\textrm{-}2& 1& \ \\
 \ & \ & \ & \ &1&\textrm{-}2&1 \\
  \ & \ & \ & \ & \ &1&\textrm{-}2
\end{bmatrix} + \begin{bmatrix}
V(x_{\textrm{-}N})& \ & \ & \  \\
 \ &\hspace{-1em} \ddots& \ & \ & \ \\
 \ & \ & & \hspace{-1em} V(x_0) \\
 \ & \ & \ & \ & \ & \hspace{-1em}\ddots \\
  \ & \ & \ & \ & \ & \ & \hspace{-1em} V(x_{N})
\end{bmatrix} \right) \begin{bmatrix}
\psi_n(x_{\textrm{-}N}) \\
\vdots \\
\psi_n(x_0) \\
\vdots \\
\psi_n(x_{N})
\end{bmatrix} = E_n \begin{bmatrix}
\psi_n(x_{\textrm{-}N}) \\
\vdots \\
\psi_n(x_0) \\
\vdots \\
\psi_n(x_{N})
\end{bmatrix} \enspace ,
\end{equation}
\end{widetext}
which takes the standard tri-diagonal form upon combining the diagonal potential term and the three-point FD approximation.

If the potential is $2 \pi$-periodic, such that the wavefunction must obey the boundary condition $\psi_n(x) = \psi_n(x \pm 2 \pi)$, we discretize this boundary condition to find:
\begin{equation}
\begin{split}
\psi_n(x_i) &= \psi_n(x_i \pm 2 \pi) \\
&= \psi_n(\Delta x \left[i  \pm 2 \pi/\Delta x \right]) \enspace ,
\end{split}
\end{equation}
where the final line indicates that $2 \pi /\Delta x$ must correspond to an integer in order to have a periodic boundary condition that connects two points in our pre-defined grid. Letting $\Delta x = 2 \pi / (2 N +1)$, we find that the terms corresponding to $\psi_n(x_{-N-1})$ and $\psi_n(x_{N+1})$ are mapped back inside the grid, i.e.,
\begin{equation}
\begin{split}
\psi_n(x_N) &= \psi_n(x_N-2 \pi) \\
&= \psi_n(\Delta x \left[ N-2N-1 \right]) \\
&=\psi_n(x_{\textrm{-}N\textrm{-}1})
\end{split}
\end{equation}
and
\begin{equation}
\begin{split}
\psi_n(x_{\textrm{-}N}) &= \psi_n(x_{\textrm{-}N}+2 \pi) \\
&= \psi_n(\Delta x \left[ -N+2N+1 \right]) \\
&= \psi_n(x_{N+1}) 
\end{split} \enspace .
\end{equation}
We then re-write the boundary term in Eq.~(\ref{eq:MATSeq}) as
\begin{widetext}
\begin{equation}
\frac{-\hbar^2}{2m} \frac{1}{\Delta x^2} \begin{bmatrix}
\psi_n(x_{\textrm{-}N\textrm{-}1}) \\
0 \\
\vdots \\
0 \\
\psi_n(x_{N+1})
\end{bmatrix} = \frac{-\hbar^2}{2m} \frac{1}{\Delta x^2} \begin{bmatrix}
\psi_n(x_N) \\
0 \\
\vdots \\
0 \\
\psi_n(x_{\textrm{-}N})
\end{bmatrix} = \frac{-\hbar^2}{2m} \frac{1}{\Delta x^2} \begin{bmatrix}
\ & \ & \ & \ & \ & \ & \ & \ &1 \\
\ & \ & \ & \ & \ & \ & \ & 0 & \ \\
\ & \ & \ & \ & \ & \ \\
\ & \ & \ & \ & \iddots & \ \\
\ & \ & \ & \ & \ & \ \\
\ & 0 & \ & \ & \ & \ \\
1 & \ & \ & \ & \ & \
\end{bmatrix} \begin{bmatrix}
\psi_n(x_{\textrm{-}N}) \\
\vdots \\
\psi_n(x_0) \\
\vdots \\
\psi_n(x_{N})
\end{bmatrix} \enspace ,
\end{equation}
and find a slightly modified eigenvalue equation to solve,
\begin{equation}
\left( \frac{-\hbar^2}{2m} \frac{1}{\Delta x^2} \begin{bmatrix}
\textrm{-}2&1& \ & \ & \ & \ &1 \\
1&\textrm{-}2&1& \ & \ \\
 \ &1&\textrm{-}2 \\
 \ & \ & &  \ddots \\
 \ & \ & \ & \ &\textrm{-}2& 1& \ \\
 \ & \ & \ & \ &1&\textrm{-}2&1 \\
  1 & \ & \ & \ & \ &1&\textrm{-}2
\end{bmatrix} + \begin{bmatrix}
V(x_{\textrm{-}N})& \ & \ & \  \\
 \ &\hspace{-1em} \ddots& \ & \ & \ \\
 \ & \ & & \hspace{-1em} V(x_0) \\
 \ & \ & \ & \ & \ & \hspace{-1em}\ddots \\
  \ & \ & \ & \ & \ & \ & \hspace{-1em} V(x_{N})
\end{bmatrix} \right) \begin{bmatrix}
\psi_n(x_{\textrm{-}N}) \\
\vdots \\
\psi_n(x_0) \\
\vdots \\
\psi_n(x_{N})
\end{bmatrix} = E_n \begin{bmatrix}
\psi_n(x_{\textrm{-}N}) \\
\vdots \\
\psi_n(x_0) \\
\vdots \\
\psi_n(x_{N})
\end{bmatrix} \enspace ,
\end{equation}
\end{widetext}
which is not strictly tri-diagonal, but still sparse.

\begin{figure}
\centering
\includegraphics[width= 0.48\textwidth]{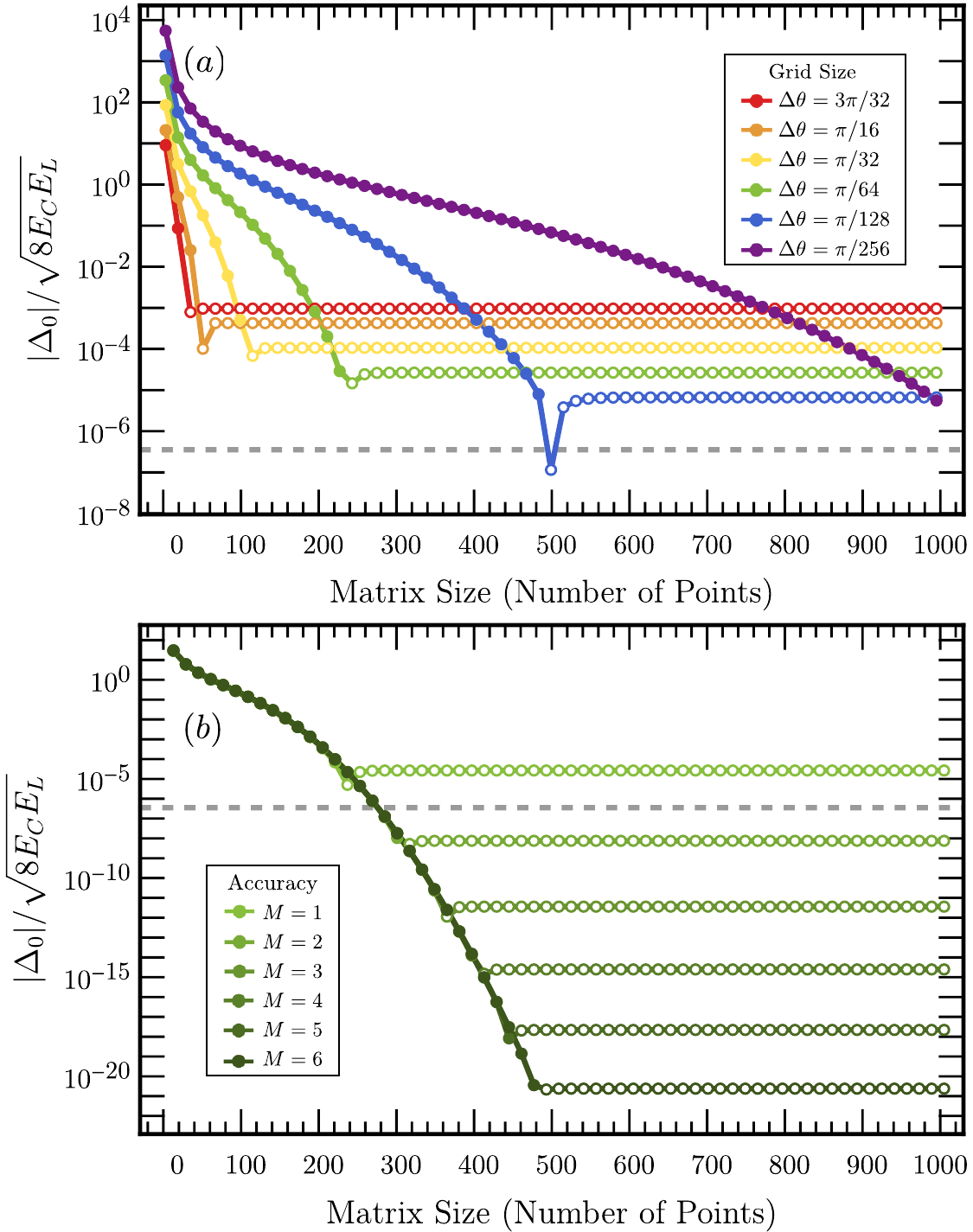}
\caption{The absolute value of the energy difference $|\Delta_0|$ as function of matrix size in the ground state of the LC oscillator. In (a), we utilize the typical, lowest-order finite difference method to compare various grid sizes $\Delta \theta$, as indicated in the included legend. In (b), we compare higher-order FD approximations in the FD method for a given grid size ($\Delta \theta = \pi/64$), indicated by the accuracy associated with the included number of points $M$. Both sets of curves are scaled according to the LC frequency. Decoherence accuracy (also scaled according to the LC frequency) is indicated with a horizontal gray dashed line. Open data points indicate where $\Delta_0 <0$, in order to note and highlight the non-variational nature of the FD approximation and method.}
\label{fig:FDLC}
\end{figure}

We highlight the implementation of the FD method in the LC oscillator, which has a bounded potential, by considering the continuous phase representation of the LC oscillator's Hamiltonian in Eq.~(\ref{eq:LCham}). We then solve an eigenvalue equation analogous to Eq.~(\ref{eq:FDbounded}) for a chosen grid size and matrix size, corresponding to the total number of points considered in the discretization. This enables the creation of Fig.~\ref{fig:FDLC}a, where we plot the absolute value of the energy difference $|\Delta_0|$ as a function of matrix size in the ground state of the LC oscillator. Fig.~\ref{fig:FDLC}a includes the curves for six grid sizes $\Delta \theta$ out of a total of seventeen that we consider, ranging from $\Delta \theta = \pi / 512$ to $\Delta \theta = 3 \pi /4$, which we use to generate the curve present in Fig.~\ref{fig:LCmetrics}b of the main text. As for each of the numerical analyses we have discussed so far, up to 150 digits of precision are maintained. Fig.~\ref{fig:FDLC}a demonstrates the overall ineffectiveness of the FD method in approximating the eigenfunctions of the LC oscillator. Decoherence accuracy is only surpassed for a single point, requiring a $499 \times 499$ matrix. Despite the increasingly smaller grid sizes, in each case, saturation is above decoherence accuracy and requires matrix sizes that are an order of magnitude larger than those needed by the DVRs we focus on in the main text.

\subsection{General derivation of the FD approximation}

We establish the discrete approximation for $\frac{\partial^2 \psi_n(x_i)}{\partial x^2}$ (and other derivatives) by using a series expansion of $\psi_n(x)$ about the grid point $x_i$, which increases in accuracy as we incorporate more of the points surrounding $x_i$ into the approximation.

Consider first the evaluation of $\psi_n(x)$ at $2M$ grid points surrounding $x_i$, such that we can define the vector $\underline{\psi_n} = \lbrace \psi_n(x_{i-M}), ...,\psi_n(x_{i}), ..., \psi_n(x_{i+M}) \rbrace$. Each of these elements can be expressed using a Taylor series expansion of $\psi_n(x)$ about the grid point $x_i$, 
\begin{equation}
\psi_n(x) = \sum_{m=0}^\infty \frac{(x-x_i)^m}{m!} \frac{\partial^m \psi_n(x_i)}{\partial x^m} \enspace ,
\end{equation} 
evaluated at the element's grid point. By truncating this series at order $2M$, we find a $(2M+1) \times (2M+1)$ matrix equation,
\begin{equation}
\underline{\psi_n} \approx \underline{\underline{F}} \ \underline{\partial \psi_n(x_i)} \enspace ,
\end{equation}
where we define the vector $\underline{\partial \psi_n(x_i)} = \lbrace \frac{\partial^0 \psi_n(x_i)}{\partial x^0} , \frac{\partial^1 \psi_n(x_i)}{\partial x^1}, ..., \frac{\partial^{2M} \psi_n(x_i)}{\partial x^{2M}}  \rbrace$ and the matrix $\underline{\underline{F}}$ as that corresponding to the terms of each Taylor series expansion. This matrix has elements $F_{jk} = \frac{\left[ \Delta x (j-M-1)\right]^{(k-1)}}{(k-1)!}$, explicitly written as
\begin{equation}
\underline{\underline{F}} = \begin{bmatrix}
1&-M\Delta x &\hdots&\hdots& \frac{(-M)^{2M}}{2M!}\Delta x^{2M}\\
\vdots&\vdots&\vdots&\vdots&\vdots  \\
1&0&\hdots&\hdots&0 \\
\vdots&\vdots&\vdots&\vdots&\vdots \\
1&M\Delta x &\hdots&\hdots& \frac{(M)^{2M}}{2M!}\Delta x^{2M}
\end{bmatrix} \enspace .
\end{equation}
Inverting the matrix $\underline{\underline{F}}$ then yields $(2M+1)$-point approximations for the derivatives in $\underline{\partial \psi_n(x_i)}$ in terms of $\psi_n(x)$ evaluated at $x_i$ and the $2M$ surrounding grid points.

For example, taking $M=1$ leads to the usual centered, three-point FD approximation for the first and second derivatives, incorporating only the point of interest and its nearest neighbors in the grid,
\begin{equation}
\frac{\partial \psi_n(x_i)}{\partial x} \approx \frac{\psi_n(x_{i+1})-\psi_n(x_{i-1})}{2 \Delta x} \enspace ,
\end{equation}
and
\begin{equation}
\frac{\partial^2 \psi_n(x_i)}{\partial x^2} \approx \frac{\psi_n(x_{i-1})-2\psi_n(x_i)+\psi_n(x_{i+1})}{\Delta x^2} \enspace .
\end{equation}
This is the approximation we utilize in the LC oscillator and in the overview of the FD method in the previous subsection.

Instead of this lowest-order approximation, we could instead incorporate any of the higher order FD derivative approximations into the FD method in the previous subsection. In that case, we must be mindful of the additional points that appear in the boundary term, either sending them to zero in the bounded case or mapping them back onto the grid in the periodic case. In the LC oscillator, its bounded potential enables a straightforward investigation of the impact of considering such higher-order approximations. In Fig.~\ref{fig:FDLC}b, we show the absolute value of the energy difference $|\Delta_0|$ as function of matrix size in the ground state of the LC oscillator, for increasingly larger accuracy $M$, all for the grid size $\Delta \theta = \pi / 64$. We do find large improvements in performance upon utilizing these higher order approximations, however, large matrices are still required. Given the interest in using this approach to provide a straightforward, well-known, and sparse point of comparison for our DVR analysis, higher-order approximations become less desirable. As the accuracy of the FD method increases, sparseness rapidly decreases, while still requiring large matrix sizes for accurate level convergence.

Of note, this symmetric, centered FD approximation has an accuracy of order $2M$, the highest derivative incorporated into each Taylor series expansion. Due to this truncation, the approximations for the derivatives and the FD method outlined in the previous subsection are non-variational, as these truncated terms can add or subtract from the exact solution, depending on the function $\psi_n(x)$. Thus, like each of the DVRs in the main text, using the FD method to model Hamiltonians and evaluate energy eigenvalues does not guarantee energies that are larger than their true value. We see evidence of this feature in Fig.~\ref{fig:FDLC}, as evidenced by the open data points and similarly sharp features in some curves.

\bibliography{DVRRefs}

\end{document}